\documentclass[journal]{IEEEtran}
\usepackage{setspace}
\usepackage{clrscode}
\usepackage{pict2e}

\usepackage{amsmath}
\usepackage{amssymb}
\usepackage{graphicx}
\usepackage{epsfig}
\usepackage{algorithm}
\usepackage{algorithmic}
\usepackage{caption}
\usepackage{amsthm}
\usepackage{subcaption}
\usepackage{dsfont}
\usepackage{mathabx}
\usepackage{stfloats}

\usepackage[ps2pdf,
bookmarks=false,
bookmarksnumbered=false, 
bookmarksopen=false, 
colorlinks=false]{}

\newcounter{TempEqCnt}
\makeatletter

\newcommand{\Rmnum}[1]{\expandafter\@slowromancap\romannumeral #1@}
\makeatother

\newcommand{\cL}{\mathcal{L}}
\newcommand{\pdf}{{\text{PDF}}}
\newcommand{\ccdf}{{\text{CCDF}}}

\newcommand{\sinr}{{\text{{SINR}}}}
\newcommand{\los}{{\text{{LOS}}}}
\newcommand{\nlos}{{\text{{NLOS}}}}

\newcommand{\Rr}{\mathbb{R}}
\newcommand{\prob}{\mathds {P}}
\newcommand{\E}{\mathbb{E}}

\newcommand{\mK}{\mathcal{K}}

\newtheorem{Lem}{Theorem}
\newtheorem{Lemm}{Lemma}

\newtheorem{Corr}{Corollary}
\newtheorem{Def}{Definition}

\allowdisplaybreaks[2]

\begin{document}

%
\title{Coverage in Downlink Heterogeneous mmWave Cellular Networks with User-Centric Small Cell Deployment}

\author{\IEEEauthorblockN{Xueyuan Wang, Esma Turgut and M. Cenk Gursoy}
\thanks{The authors are with the Department of Electrical
Engineering and Computer Science, Syracuse University, Syracuse, NY, 13244
(e-mail: xwang173@syr.edu, eturgut@syr.edu, mcgursoy@syr.edu).}}

\maketitle

\begin{abstract}
A $K$-tier heterogeneous downlink millimeter wave (mmWave) cellular network with user-centric small cell deployments is studied in this paper. In particular, we consider a heterogeneous network model with user equipments (UEs) being distributed according to a Poisson Cluster Process (PCP). Specifically, we address two cluster processes, namely (i) Thomas cluster process, where the UEs are clustered around the base stations (BSs) and the distances between UEs and the BS are modeled as Gaussian distributed, and (ii) Mat\'ern cluster process, where the UEs are scattered according to a uniform distribution. In addition, distinguishing features of mmWave communications including directional beamforming and a sophisticated path loss model incorporating both line-of-sight (LOS) and non-line-of-sight (NLOS) transmissions, are taken into account. Initially, the complementary cumulative distribution function (CCDF) and probability density function (PDF) of path loss are provided. Subsequently, using tools from stochastic geometry, we derive a general expression for the signal-to-interference-plus-noise ratio (SINR) coverage probability. Our results demonstrate that coverage probability can be improved by decreasing the size of UE clusters around BSs, decreasing the beamwidth of the main lobe, or increasing the main lobe directivity gain. Moreover, interference has noticeable influence on the coverage performance of our model. We also show that better coverage performance is achieved in the presence of clustered users compared to the case in which the users are distributed according to a Poisson Point Process (PPP).
\end{abstract}

\thispagestyle{empty}

\section{Introduction}
 Demand for cellular data has been growing rapidly in recent years resulting in a global bandwidth shortage for wireless service providers \cite{Milli_TSP, Milli_SunR}. In the presence of this severe spectrum shortage in conventional cellular bands, millimeter wave (mmWave) frequencies between 30 and 300 GHz have been attracting growing attention for deployment in next-generation wireless heterogeneous networks \cite{Milli_MRA}. Larger bandwidths available in mmWave frequency bands make them attractive to meet the exponentially growing demand in data traffic \cite{Milli_Yao}. On the other hand, communication in mmWave frequency bands has several limitations such as increase in free-space path loss with increasing frequency and poor penetration through solid materials. However, with the use of large antenna arrays by utilizing the shorter wavelengths of mmWave frequency bands, and enabling beamforming at the transmitter and receiver, frequency dependent path-loss can be compensated \cite{Milli_ZhouP}. Additionally, with the employment of directional antennas, out-of-cell interference can be reduced greatly.

Future mobile networks are converging towards being heterogeneous, i.e., supporting the coexistence of denser but lower-power small-cell base stations (BSs) with the conventional high-power and low-density large-cell BSs \cite{cluster_Yuang} \cite{Milli_HSDhillon} \cite{Milli_HSJo}. Heterogeneous mmWave cellular networks have been addressed in several recent studies. An energy-efficient mmWave backhauling scheme for small cells in 5G is considered in \cite{Milli_YNiu}, where the small cells are densely deployed and a macrocell is coupled with small cells to some extent. Mobile users are associated with BSs of the small cells, and have the communication modes of both fourth-generation access and mmWave backhauling operation. The macrocell BS and small-cell BSs are also equipped with directional antennas both for 4G communications and transmissions in the mmWave band. A general multi-tier mmWave cellular network is studied in \cite{Milli_RenzoM} and \cite{Milli_Esma}. The BSs in each tier are distributed according to a homogeneous Poisson point process (PPP) with certain densities. Moreover, in \cite{Milli_RenzoM} a two-ball approximation is considered, modeling the state of links in line of sight (LOS), non-LOS (NLOS), and outage. In \cite{Milli_Esma}, a $K$-tier heterogeneous mmWave cellular network is considered, and signal-to-interference-plus-noise-ratio (SINR) coverage probability is derived by incorporating the distinguishing features of mmWave communications, and a $D$-ball  approximation for blockage modeling is employed. In \cite{Esma-VTC-F17}, we have analyzed the uplink performance of device-to-device (D2D)-enabled mmWave cellular networks. However, UEs are located independently with BS locations in these works.

Stochastic geometry has become a powerful tool for analyzing cellular networks in recent years. As also noted above, a common approach is to model the locations of BSs and user equipments (UEs) randomly and independently using the PPP distribution. However, assuming BS and UE locations independent from each other is not quite accurate. In practice, UE density is expected to be higher around some low-power small cell BSs causing a correlation in the locations of BSs and UEs. Therefore, user-centric deployment of small cells is becoming an important part of future wireless architectures \cite{cluster_FB}. In this type of deployment, UEs are considered to be clustered around the small-cell BS which is considered as the cluster center.

Several recent studies have also attempted to model the UEs as clustered around the small-cell BSs. In \cite{cluster_YZhong}, the authors consider Neyman-Scott cluster process, in which the centers of the clusters and cluster members are assumed to be distributed according to some stationary PPP independent from each other. Although the cluster process is considered, the correlation between the locations of the cluster centers and members is not taken into account. In \cite{cluster_CChen}, PPP-Poisson cluster process (PCP) model is employed in which macrocell BS locations are modeled according to a PPP, while picocell BS locations are distributed according to a PCP. Authors investigate the the effect of the distance between the BS and UEs on coverage probability. In \cite{cluster_HTaba}, a multi-cell uplink non-orthogonal multiple access  system is provided. BSs are distributed according to a homogeneous PPP, and UEs are uniformly clustered around the BSs within a circular region. Three scenarios are considered in \cite{cluster_HTaba}, including perfect  successive interference cancellation (SIC), imperfect SIC and imperfect worst case SIC at the receiver side. Moreover, the Laplace transform of the interference is analyzed. In \cite{cluster_CSaha}, authors consider a $K$-tier heterogeneous network (HetNet) model with user-centric small cell deployments in which the locations of UEs are modeled by a PCP with one small cell BS located at the center of each cluster process. They also specialize the PCP as a Thomas cluster process where the UEs are Gaussian distributed around the small BSs, and a Mat\'ern cluster process where the UEs are uniformly distributed inside a disc centered around the location of small cell BSs. In addition to modeling locations of UEs  as a PCP, small-cell BS clustering is considered in \cite{cluster_MAf}
 to capture the correlation between the large-cell and small-cell BS locations. A unified HetNet model in which a fraction of UEs and some BS tiers are modeled as PCPs is developed in \cite{cluster_CSahaPCP} to reduce the gap between the real-word deployments and the popular PPP-based analytical model. However, these prior studies that considered clustered users have not addressed transmission in mmWave frequency bands.

In this paper\footnote{A short conference version of this paper has been submitted to the 2017 PIMRC, Montreal, Canada \cite{Xueyuan-Esma-PIMRC17}.}, motivated by the facts that mmWave is poised to be an important component of next generation wireless networks and clustered UEs are experienced in several practical scenarios, we analyze a $K$-tier heterogeneous downlink mmWave cellular network with UE-centric small cell deployments. Our main contributions can be summarized as follows:

\begin{itemize}

\item We develop a new and more practical heterogeneous mmWave cellular network model by considering the correlation between the locations of UEs and BSs. In particular, Thomas cluster processes and Mat\'ern cluster processes are considered to model the locations of UEs around the small-cell BSs.

\item Cell association probabilities are determined by deriving the complementary cumulative distribution function (CCDF) and probability distribution function (PDF) of the path loss for each tier by employing averaged biased-received power cell association criterion.

\item A general expression for SINR coverage probability is obtained by considering PCP distributed UEs and incorporating the distinguishing features of mmWave communication such as directional beamforming and having different path loss laws for LOS and NLOS links. $D$-ball approximation is employed for blockage modeling.

\end{itemize}

The rest of the paper is organized as follows. In Section II, we introduce the system model. CCDF and PDF of the path loss, and association probabilities for each tier are derived in Section III. In Section IV, the total SINR coverage probability of the entire network is obtained. In Section V, numerical and simulation results are presented to investigate the impact of several system parameters on the coverage probability performance. Finally, the conclusions are drawn and future work is discussed in Section VI. Proofs are relegated to the Appendix.

\section{System Model} \label{SModel}
\subsection{Base Station Distribution Modeling}
In our model, a $K$-tier heterogeneous downlink mmWave cellular network is considered. BSs in all tiers are distributed according to a homogeneous PPP (more specifically, the BSs in the $j^{th}$ tier are distributed according to PPP $\Phi_j$ of density $\lambda_j$ on the Euclidean plane for $j \in \mK = \{1, 2, ... , K\}$), and are assumed to be transmitting in a mmWave frequency band. BSs in the $j^{th}$  tier are distinguished by their transmit power $P_j$, biasing factor $B_j$, and blockage model parameters.

\begin{figure*}[htbp]
\centering
\begin{minipage}{4cm}
\centering
\includegraphics[width=5.5cm]{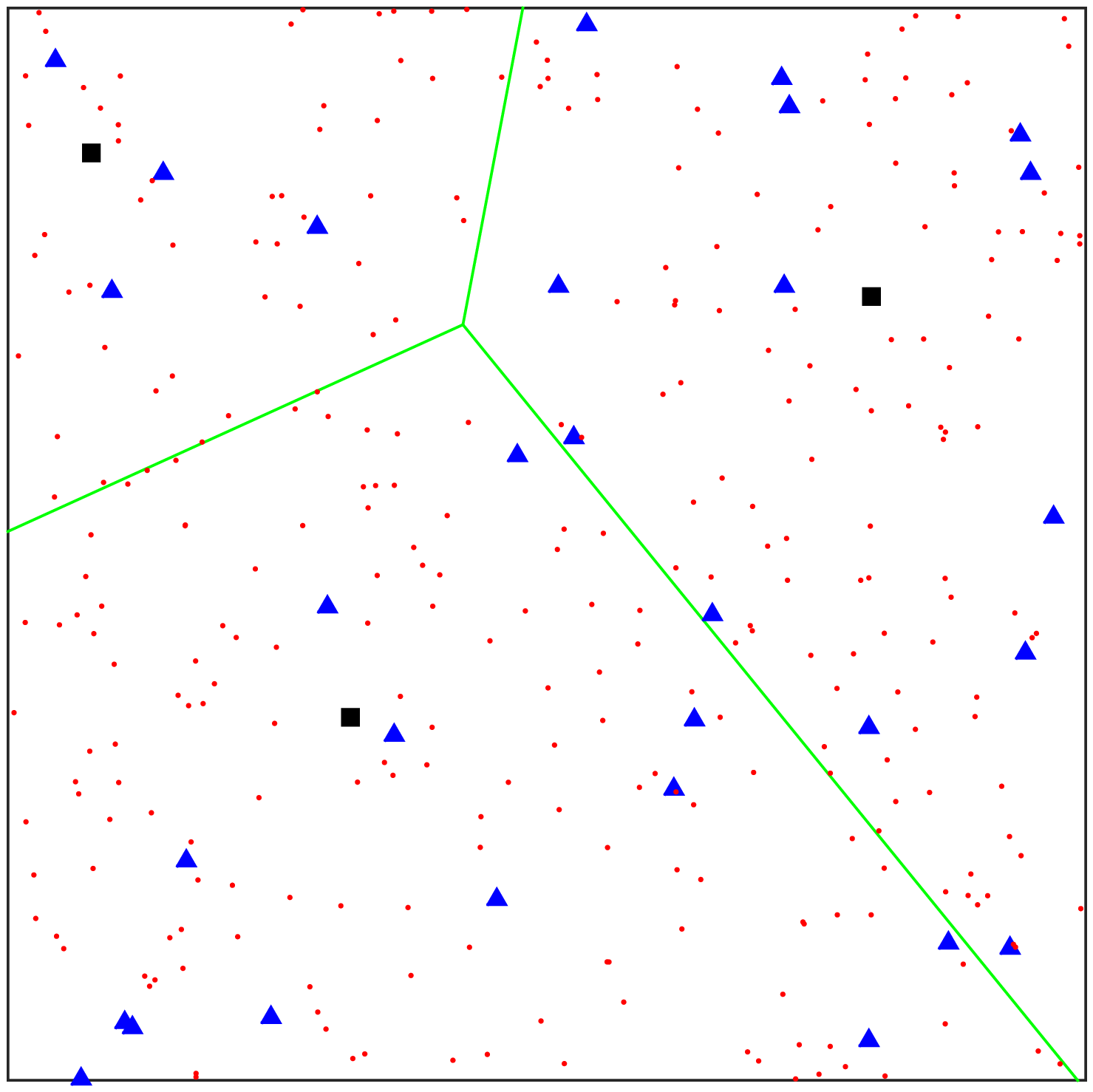}
\subcaption{\scriptsize Users are uniformly distributed. }
\end{minipage}
\hfill
\begin{minipage}{4cm}
\centering
\label{fig:subfig:b}
\includegraphics[width=5.5cm]{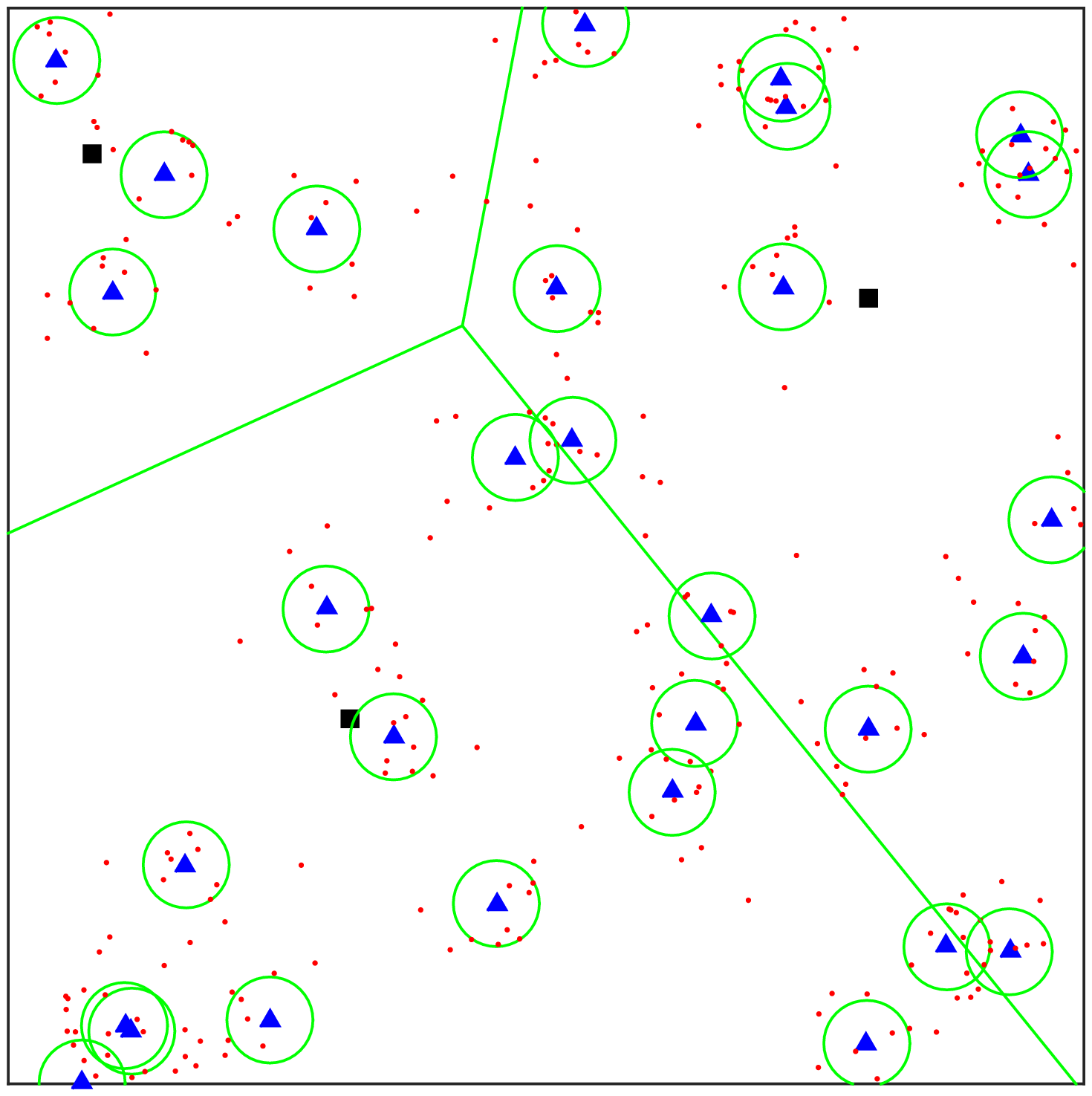}
\subcaption{\scriptsize Users in Thomas cluster process. }
\end{minipage}
\hfill
\begin{minipage}{5cm}
\centering
\includegraphics[width=5.5cm]{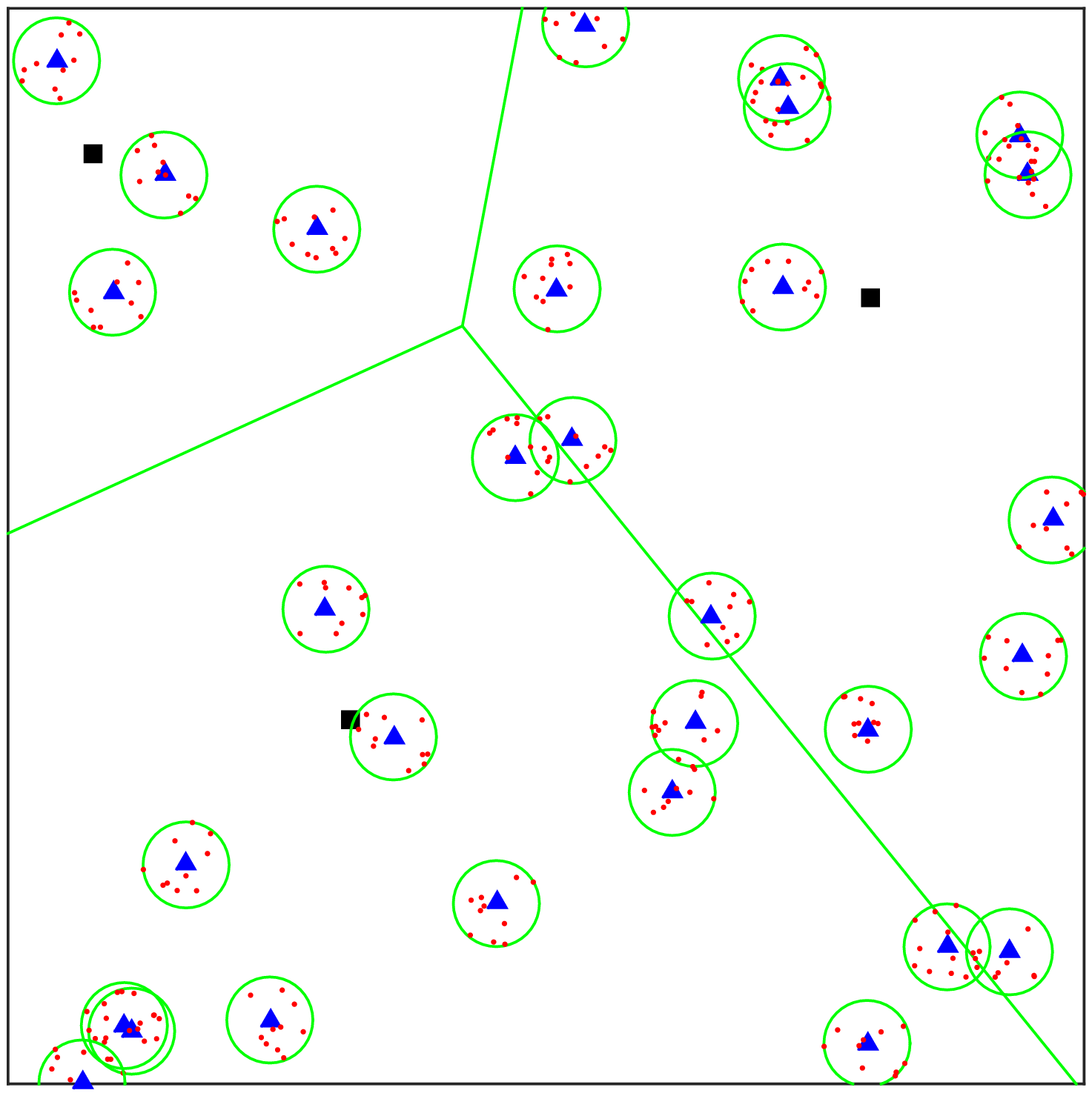}
\subcaption{\scriptsize Users in Mat\'ern cluster process.}
\end{minipage}
\caption{\small Two-tier heterogeneous network model, where microcells (black squares) and picocells (blue triangles) are distributed as independent PPPs. (a) UEs are uniformly and independently distributed. (b) UEs are distributed around picocells according to a Gaussian distribution. (c) UEs are distributed around picocells according to a uniform distribution. The average number of UEs per cluster is 10 in (b) and (c). \normalsize}
\end{figure*}
\subsection{User Distribution Modeling}
Unlike previous works which mostly consider UEs distributed uniformly according to some independent homogeneous point process, we consider a more realistic network scenario where the UEs are clustered around the smaller cell BSs. In this network scenario, smaller cell BSs are located at the center of the clustered UEs where the locations of the UEs are modeled as a PCP. UEs in each cluster are called cluster members. The cluster where the typical UE comes from is called the representative cluster.

Cluster members are  assumed to be symmetrically independently and identically distributed (i.i.d.) around the cluster center. Assume that the cluster center is a BS in the $j^{th}$ tier, then the union of cluster members' locations form a PCP, denoted by $\Phi_u^j$. In this paper, $\Phi_u^j$ is modeled as either (i) a Thomas cluster process or (ii) a Mat\'{e}rn cluster process. If a Thomas cluster process is considered, the UEs are scattered according to a Gaussian distribution with variance $\sigma_j^2$. If UEs' locations are denoted as $\textbf{Z}_u^{j} \in \Rr^2$ with respect to its cluster center, then the PDF of the distance is given by \cite{clusterbook_HMartin}
\begin{align}
f_{\textbf{Z}_u^{j}} (\textbf{z}) = \frac{1}{2 \pi \sigma_{j}^{2}} \exp (-\frac{||\textbf{z}||^2}{2 \sigma_j^2}) \qquad \textbf{z} \in \Rr^2.
\end{align}
If a Mat\'ern cluster process is considered, then the UEs are scattered according to a uniform distribution, i.e., UEs are symmetrically uniformly spatially distributed around the cluster center within a circular disc of radius $R_j$ and thus the PDF of the distance is
\begin{align}
f_{\textbf{Z}_u^{j}} (\textbf{z}) = \frac{1}{\pi R^2_j} \qquad ||\textbf{z} || \leq R_j
\end{align}
where $\textbf{z}$ is the realization of the random vector $\textbf{Z}_u^{j}$ in Cartesian domain. A two-tier heterogeneous network model with different UE distributions is shown in Fig. 1, where microcells have relatively higher power and picocells have relatively lower power but larger density. While the UEs are distributed according to a homogeneous PPP in Fig. 1(a), they follow Thomas cluster and Mat\'{e}rn cluster processes around the picocell BSs in Fig. 1(b) and Fig. 1(c), respectively.

Without loss of generality, the typical UE is assumed to be located at the origin. Therefore, $\textbf{Y}_0$, denoting the relative location of the cluster center with respect to the typical UE, has the same distribution as $\textbf{Z}_u^{j}$. Next, we transfer $\textbf{Y}_0(t_1,t_2)$ from Cartesian coordinates to polar coordinates $(Y_0,\Theta)$, using standard transformation techniques as follows:
\begin{align}
f_{Y_0,\Theta}(y_0,\theta)=f_{y_0}(t_1,t_2) \times \bigg| \partial \bigg(\frac{t_1,t_2}{y_0,\theta}  \bigg)\bigg| ,
\end{align}
where
\begin{align}
 \partial \bigg(\frac{t_1,t_2}{y_0,\theta}  \bigg) =
 \bigg[
 \begin{array}{ccc}
  \frac{ \partial t_1}{\partial y_0}  &\frac{ \partial t_1}{\partial \theta} \\
 \frac{ \partial t_2}{\partial y_0}   &\frac{ \partial t_2}{\partial \theta}
 \end{array}
 \bigg] \notag .
\end{align}
Marginal distribution of the distance $Y_0$ can be obtained from the joint distribution by integrating over $\theta$ as follows:
\begin{align}
f_{Y_0}(y_0) = \int_0^{2\pi} f_{Y_0,\Theta}(y_0,\theta) d\theta .
\end{align}
Therefore, (i) if $\Phi_u^j$ is a Thomas cluster process,  the CCDF and PDF of $Y_0$ are given as \cite{cluster_MAf_Model}
\begin{align}
&\text{CCDF: }\qquad \overline{F}_{Y_0}(y_0)=\exp \left( \frac{-{y^2_0}}{2{\sigma^2_j}}\right)   \qquad (y_0 \geq 0 ), \\
&\text{PDF:    }\qquad f_{Y_0}(y_0)=\frac{y_0}{{\sigma^2_j}}\exp\left(\frac{-{y^2_0}}{2{\sigma^2_j}}\right)  \qquad (y_0 \geq 0 ),
\end{align}
where $\sigma_j^2$ is the variance of the distance between the typical UE and cluster center; (ii) if $\Phi_u^j$ is a Mat\'ern cluster process,  the CCDF and PDF of $Y_0$ are given as
\begin{align}
&\ccdf :  \qquad \overline{F}_{Y_0}(y_0)=1-\frac{{y_0}^2}{R^2_j} \qquad       (0\leq{y_0}\leq{R_j}) \\
&\pdf : \qquad f_{Y_0}(y_0)=\frac{2{y_0}}{R^2_j}     \qquad         (0\leq{y_0}\leq{R_j})
\end{align}
where $R_j$ is the radius of the representative cluster in the $j^{th}$ tier.

Note that BSs in the $j^{th}$ tier are distributed according to a PPP $\Phi_j $ $( j\in \mK)$ and the typical UE is assumed to be served by the nearest BS in the $j^{th}$ tier. Let $y_j$ denote the distance  from the typical UE to the nearest BS in the $j^{th}$ tier. Then, the CCDF and PDF of  $y_j$ are given as \cite{clusterbook_HMartin}
\begin{align}
&\text{CCDF: }\qquad \overline{F}_{Y_j}(y_j)=\exp(-\pi{\lambda_j}{y^2_j})       \quad (y_j \geq 0 ),  \\
&\text{PDF: }\qquad f_{Y_j}(y_j)=2\pi{\lambda_j}{y_j}\exp(-\pi{\lambda_j}{y^2_j})   \quad (y_j \geq 0 ),
\end{align}
where  $\lambda_j$ is the density of PPP $\Phi_j$.

Similar to \cite{cluster_CSaha}, for notational simplicity, we form an additional tier, named as $0^{th}$ tier, which includes the cluster center of the typical UE. Thus, our model is denoted as a $\mK_1 = \{ 0\} \cup \mK = \{0,1,2,...,K\} $ tier model.

\subsection{Antenna and Channel Modeling}
In this setting, we have the following assumptions regarding the antenna and channel models of the $K$-tier heterogeneous downlink mmWave cellular network:

\subsubsection{Directional beamforming}
Antenna arrays at all BSs and UEs are assumed to perform directional beamforming. For analytical tractability, sectored antenna model is employed where $M$, $m$, $\theta$ denote the main lobe directivity gain, side lobe gain and beamwidth of the main lobe, respectively \cite{Milli_Esma, Milli_Bai}.
We assume perfect beam alignment between the typical UE and  its serving BS resulting in a overall antenna gain of $MM$. In other words, the typical UE and its serving BS can adjust their antenna steering orientation using the estimated angles of arrivals to achieve maximum directivity gain. Beam direction of the interfering links is modeled as a uniform random variable on [0, 2$\pi$]. Hence, the effective antenna gain $G$ between the typical UE and an interfering BS can be described with the following random variable:
\begin{align}
&G=
\begin{cases}
MM   \qquad  \text{with probability } P_{MM} =(\frac{\theta}{2\pi})^2  \\
Mm     \qquad  \text{with probability } P_{Mm} =2\frac{\theta}{2\pi}  \frac{2\pi -\theta}{2 \pi}  \\
mm     \qquad  \text{ with probability } P_{MM} =(\frac{2\pi - \theta}{2\pi})^2 ,
\end{cases}
\end{align}
where $M$ is the main lobe directivity gain, $m$ is the side lobe gain, $\theta$ is the beamwidth of the main lobe, and $p_G$ is the probability of having the antenna gain of $G \in \{MM, Mm, mm\}$.

\begin{figure}[!h]
\centering
  \includegraphics[width=0.35\textwidth]{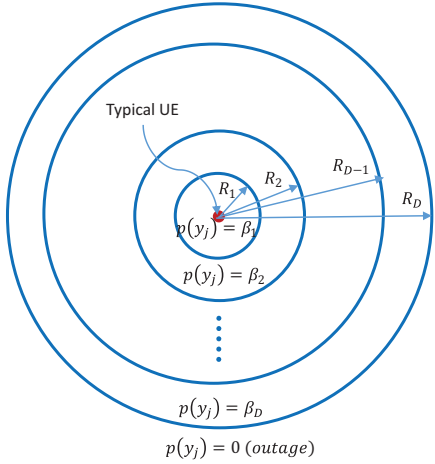}
  \caption{LOS ball model}
\label{D-ball}
\end{figure}
\subsubsection{Path loss and blockage modeling}
Link between a typical UE and a BS can be either a LOS or NLOS link. A LOS link occurs when there is no blockage between the UE and  the BS, while a NLOS link occurs between the UE and the BS if blockage exists. An additional outage state can occur if the path-loss is sufficiently high causing no link establishment between the UE and the BS \cite{Milli_RenzoM}.

Consider an arbitrary link of length $y_j$ ($j \in \mK$), and define the LOS probability function $p(y_j)$ as the probability that the link is LOS. In \cite{Milli_RenzoM} and \cite{Milli_DingM}, authors employ multi-ball models with piece-wise LOS probability functions. Similar to the piece-wise LOS probability function approach, $D$-ball approximation model is adopted in \cite{Milli_Esma}. In this paper, we employ the same $D$-ball approximation model used in \cite{Milli_Esma}. As shown in Fig. 2, a link is in LOS state with probability $p(y_j)=\beta_{j1}$ inside the first ball with radius $R_1$, while NLOS state occurs with probability $1-\beta_{j1}$. Similarly, LOS probability is equal to $p(y_j)=\beta_{jd}$ for $y_j$ between $R_{d-1}$ and $R_d$ for $d=2,\ldots,D$, and all links with distances greater than $R_D$ are assumed to be in outage state. Additionally, LOS and NLOS links have different path loss exponents in different ball layers. Therefore, the path loss on each link in the $j^{th}$ tier $(j \in \mK)$ can be expressed as follows:

\begin{align}
\hspace{0 cm}&L_{j}(y_j) =  \notag \\
&\begin{cases}
          \begin{cases}
           \kappa_1^L{y_j}^{\alpha_1^{jL}}  \text {with  prob. }\beta_{j1}    \\
           \kappa_1^N{y_j}^{\alpha_1^{jN}}  \text {with  prob. } (1-\beta_{j1})
          \end{cases} \hspace{0.3cm} \text{if}\quad r\leq R_{j1}\\
          \begin{cases}
           \kappa_2^L{y_j}^{\alpha_2^{jL}}  \text {with  prob. }\beta_{j2}    \\
         \kappa_2^N{y_j}^{\alpha_2^{jN}}  \text {with  prob. } (1-\beta_{j2})
          \end{cases} \hspace{0.3cm} \text{if}\quad R_{j1}\leq r\leq R_{j2}\\
           \quad  \vdots \\
          \begin{cases}
            \kappa_D^L{y_j}^{\alpha_D^{jL}}  \text {with  prob. }\beta_{jD}    \\
             \kappa_D^N{y_j}^{\alpha_D^{jN}}  \text {with  prob. } (1-\beta_{jD})
          \end{cases}\text{if }R_{j(D-1)}\leq r\leq R_{jD}\\
            \text{outage \quad if } r\geq R_{jD},
\end{cases}
\end{align}
where $\alpha_d^{jL}, \alpha_d^{jN }$ are the LOS and NLOS path loss exponents, respectively, for the $d^{th}$ ball of the $j^{th}$ tier, $\kappa_d^L$, $\kappa_d^N$ are the path loss of LOS and NLOS links at a distance of 1 meter in the $d^{th}$ ball, respectively, and $R_{jd}$ is the radius for $d^{th}$ ball in the $j^{th}$ tier $(j\in \mK), \text{ for } d= 1,2,...,D$.

For the $0^{th}$ tier, since there is only one BS which is at the cluster center and the distance between the cluster center and UE is relatively small, $1$-ball model is employed with no outage being considered. Therefore, the path loss of the link in the $0^{th}$ tier can be expressed as follows:
\begin{align}
\hspace{0 cm}L_{0}(y_0) &=
\begin{cases}
           \kappa_1^L{y_0}^{\alpha_1^{0L}}  \text {with  prob. }\beta_{01}    \\
           \kappa_1^N{y_0}^{\alpha_1^{0N}}  \text {with  prob. } (1-\beta_{01}),
\end{cases}
\end{align}
where similar notations are used for path loss parameters.

A summary of notations is provided in Table I.

\begin{table}[htbp]
\caption{Notations Table}
\centering
\begin{tabular}{|l | l |}
\hline
\small \textbf{Notations} &  \small \textbf{Description}  \\  \hline
 \scriptsize$\Phi_j, \lambda_j $& \scriptsize PPP of BSs of the $j^{th}$ tier, the density if $\Phi_j$  \\ \hline
\scriptsize$\Phi_u^j$  &  \scriptsize  PCP of UEs of the $j^{th}$ tier, the variance of $\Phi_u^j$\\\hline
\scriptsize$\sigma^2_j$  &  \scriptsize  The variance of $\Phi_u^j$, if $\Phi_u^j$ is a Thomas cluster process\\\hline
\scriptsize$R_j$  &  \scriptsize  The cluster size of the $j^{th}$ tier, if $\Phi_u^j$ is a Mat\'ern cluster process \\ \hline
\scriptsize$P_j,B_j$  &\scriptsize   The transmit power and biasing factor of BSs in the $j^{th}$ tier \\ \hline
\scriptsize$M,m$  &\scriptsize  The main lobe directivily gain, side lobe gain\\ \hline
\scriptsize$\theta$  &\scriptsize  Beamwidth of the main lobe\\ \hline
\scriptsize$G$  &\scriptsize  The effective antenna gian\\ \hline
\scriptsize$R_{jd}$ & \scriptsize  The size of the $d^{th}$ ball of the $j^{th}$ tier          \\ \hline
\scriptsize$\beta_{jd}$ & \scriptsize  The probability of a LOS link in the $d^{th}$ ball of the $j^{th}$ tier          \\ \hline
\scriptsize$\alpha^s_j$ & \scriptsize The path loss exponent of a LOS/NLOS link of the $j^{th}$ tier \\ \hline
\scriptsize$\kappa^s_d$ & \scriptsize The path loss of a LOS/NLOS link at a distance of 1 meter \\
 & \scriptsize in the $d^{th}$ ball \\ \hline
\scriptsize$y_j$ &\scriptsize The distance from the typical UE to the BSs in the $j^{th}$ tier \\ \hline
\scriptsize$l_j$  &\scriptsize The path loss to a BS at distance $y_j$ in the $j^{th}$ tier \\ \hline
\scriptsize$l_{j,s}$  &\scriptsize The path loss to a LOS/NLOS BS at distance $y_j$ in the $j^{th}$ tier\\ \hline
\scriptsize$h_j, \sigma^2_{n,j}$   &\scriptsize The Rayleigh gain, the noise factor \\ \hline
\end{tabular}
\end{table}

\section{Association Probability} \label{AP}
In this section,  first the CCDF and the PDF of the path loss for all tiers are determined. Subsequently, association probability is defined and formulated.

\subsection{CCDF and PDF of Path Loss in the $0^{th}$ tier}

\begin{Lemm}
The CCDF and PDF of the path loss from a typical UE to the BS in the $0^{th}$ tier can be formulated as follows:

(i) If $\Phi_u^j$ is a Thomas cluster process, then
\begin{align}
&\ccdf : \notag \\
&\overline{F}_{L_{0}}(x) = \sum\limits_{s \in \{\los,\nlos\}} \prob_{L_{0,s}} \exp\left(-\frac{1}{2{\sigma_j}^2}\left(\frac{x}{\kappa_1^s}\right)^\frac{2}{\alpha_1^{0s}}\right)  \notag \\ &\hspace{2.6in} (x \geq 0 ),  \\
&\pdf: \notag \\
&\hspace{-.5cm}f_{L_{0}}(x)=\sum\limits_{s \in \{\los,\nlos\}} \prob_{L_{0,s}}  \frac{{x}^{{\frac{2}{\alpha_1^{0s}}}-1}}{\alpha_1^{0s}{\kappa_1^s}^{\frac{2}{\alpha_1^{0s}}}{\sigma^2_j}}\exp\left(-\frac{1}{2{\sigma^2_j}}\left(\frac{x}{\kappa_1^s}\right)^\frac{2}{\alpha_1^{0s}}\right) \notag \\ &\hspace{2.6in}  (x \geq 0 )
\end{align}
where $\prob_{L_{0,\los}}= \beta_{01}$, $\prob_{L_{0,\nlos}}=1- \beta_{01}$, and $\sigma_j^2$ is the variance of UE distribution.

(ii) If $\Phi_u^j$ is a Mat\'ern cluster process, then
\begin{align}
&\ccdf : \notag \\
&\overline{F}_{L_{0}}(x) = \sum\limits_{s \in \{\los,\nlos\}} \prob_{L_{0,s}}  \left( 1-\frac{{l_{0,s}}^{\frac{2}{\alpha_1^{ks}}}}{{\kappa_1^s}^{\frac{2}{\alpha_1^{ks}}}R^2_j}  \right)    \notag \\ &\hspace{1.9in}  (0\leq l_{0,s} \leq \kappa_1^s R_j^ { \alpha_1^{ks} }), \\
&\pdf: \notag \\
&\hspace{0 cm}f_{L_{0}}(x)=\sum\limits_{s \in \{\los,\nlos\}} \frac{2 \prob_{L_{0,s}}  l_{0,s}^{{\frac{2}{\alpha_1^{ks}}}-1}}{\alpha_1^{ks}{\kappa_1^s}^{\frac{2}{\alpha_1^{ks}}}R^2_j}    \notag \\ & \hspace{1.9in} (0\leq l_{0,s} \leq \kappa_1^s R_j^ { \alpha_1^{ks} }) ,
\end{align}
where $R_j$ is the radius of the representative cluster.

Proof: See Appendix A.
\end{Lemm}

Also, the CCDF and PDF of the path loss from a typical UE to the LOS/NLOS BS in the $0^{th}$ tier can be expressed as follows:

(i) If $\Phi_u^j$ is a Thomas cluster process,
\begin{align}
&\ccdf : \notag \\
&\overline{F}_{L_{0,s}}(x)=\exp \left(-\frac{1}{2{\sigma^2_j}}\left(\frac{x}{\kappa_1^s}\right)^\frac{2}{\alpha_1^{0s}}\right)    \qquad (x \geq 0 ), \\
&\pdf: \notag \\
&\hspace{0cm}f_{L_{0,s}}(x)=\frac{{x}^{{\frac{2}{\alpha_1^{0s}}}-1}}{\alpha_1^{0s}{\kappa_1^s}^{\frac{2}{\alpha_1^{0s}}}{\sigma^2_j}}\exp\left(-\frac{1}{2{\sigma^2_j}}\left(\frac{x}{\kappa_1^s}\right)^\frac{2}{\alpha_1^{0s}}\right)    \quad (x \geq 0 )
\end{align}

(ii) If $\Phi_u^j$ is a Mat\'ern cluster process,
\begin{align}
&\ccdf : \notag \\
& \overline{F}_{L_{0,s}}(l_{0,s})=1-\frac{{l_{0,s}}^{\frac{2}{\alpha_1^{ks}}}}{{\kappa_1^s}^{\frac{2}{\alpha_1^{ks}}}R^2_j}    \qquad         (0\leq l_{0,s} \leq \kappa_1^s R_j^ { \alpha_1^{ks} }) \\
&\pdf :  \notag \\
&f_{L_{0,s}}(l_{0,s})=\frac{2{l_{0,s}}^{{\frac{2}{\alpha_1^{ks}}}-1}}{\alpha_1^{ks}{\kappa_1^s}^{\frac{2}{\alpha_1^{ks}}}R^2_j}   \qquad         (0\leq l_{0,s} \leq \kappa_1^s R_j^ { \alpha_1^{ks} })
\end{align}
where $s\in \{ \los, \nlos\}$.

\setcounter{TempEqCnt}{\value{equation}}
\setcounter{equation}{22}
\begin{figure*}
\begin{align}
\hspace{-1cm}\Lambda_j([0,x))=\pi\lambda_j \sum\limits_{d=1}^D\bigg\{\beta_{jd} \Big[ (R_{jd}^2-R_{j(d-1)}^2)\mathds{1}(x>\kappa_d^LR_{jd}^{\alpha_d^{jL}})+((\frac{x}{\kappa_d^L})^{\frac{2}{\alpha_d^{jL}}}-{R_{j(d-1)}}^2)\mathds{1} (\kappa_d^L {R_{j(d-1)}^{\alpha_d^{jL}}} <x< \kappa_d^L {R_{jd}^{\alpha_d^{jL}}} ) \Big] \notag\\
+ (1- \beta_{jd} )\Big[ (R_{jd}^2-R_{j(d-1)}^2)\mathds{1}(x>\kappa_d^NR_{jd}^{\alpha_d^{jN}})+((\frac{x}{\kappa_d^N})^{\frac{2}{\alpha_d^{jN}}}-{R_{j(d-1)}}^2)\mathds{1} (\kappa_d^N {R_{j(d-1)}^{\alpha_d^{jN}}} <x< \kappa_d^N {R_{jd}^{\alpha_d^{jN}}} ) \Big]
\bigg \}. \\
\hspace{-1cm}\Lambda_{j,\text{LOS}}([0,x))=\pi\lambda_j \sum\limits_{d=1}^D\beta_{jd} \Big[ (R_{jd}^2-R_{j(d-1)}^2)\mathds{1}(x>\kappa_d^LR_{jd}^{\alpha_d^{jL}})+((\frac{x}{\kappa_d^L})^{\frac{2}{\alpha_d^{jL}}}-{R_{j(d-1)}}^2)\mathds{1} (\kappa_d^L {R_{j(d-1)}^{\alpha_d^{jL}}} <x< \kappa_d^L {R_{jd}^{\alpha_d^{jL}}} ) \Big], \\
\hspace{-1cm}\Lambda_{j,\text{NLOS}}([0,x))=\pi\lambda_j \sum\limits_{d=1}^D(1- \beta_{jd} )\Big[ (R_{jd}^2-R_{j(d-1)}^2)\mathds{1}(x>\kappa_d^NR_{jd}^{\alpha_d^{jN}})+((\frac{x}{\kappa_d^N})^{\frac{2}{\alpha_d^{jN}}}-{R_{j(d-1)}}^2)\mathds{1} (\kappa_d^N {R_{j(d-1)}^{\alpha_d^{jN}}} <x< \kappa_d^N {R_{jd}^{\alpha_d^{jN}}} ) \Big].
 \end{align}
  \hrulefill
 \end{figure*}
 \setcounter{equation}{\value{TempEqCnt}}

\subsection{CCDF and PDF of Path Loss in the $j^{th}$ tier ($j \in \mK$) }

The following characterizations on the CCDF and PDF of path loss have been determined in \cite{Milli_Esma} (where no user clustering is considered).

\begin{Lemm}\cite[Appendix A]{Milli_Esma}
The CCDF of the path loss from a typical UE to the BS in the $j^{th}$ tier can be formulated as
\begin{align}
\overline{F}_{L_j}(x)=\exp(-\Lambda_j([0,x))) \qquad \text{for  } j \in \mK,
\end{align}
where $\Lambda_j([0,x))$ is given in (23) at the top of the next page, and  $\mathds{1}(\cdot)$ is the indicator function. Also note that $R_{j0} = 0$.
\end{Lemm}

\begin{Corr}\cite[Lemma 2]{Milli_Esma}
The CCDF of the path loss from the typical UE to the LOS/NLOS BS in the $j$th tier can be formulated as
\setcounter{equation}{25}
\begin {align}
\overline{F}_{L_{j,s}}(x)=\exp(-\Lambda_{j,s}([0,x))) \qquad \text{for  } j\in \mK,
\end{align}
where $s \in \{\los,\nlos\}$ and $\Lambda_{j,s}([0,x))$ is defined for LOS and NLOS, respectively, as in (24) and (25) given at the top of the next page.

\end{Corr}

 Also, the PDF of $L_{j,s}(y)$, denoted by $f_{L_{j,s}}$, which will be used in the following section, is given by
\begin {align}
&f_{L_{j,s}}(x)=-\frac{d \overline{F}_{L_{j,s}}(x)}{dx} = \Lambda'_{j,s}([0,x))\exp(-\Lambda_{j,s}([0,x))) \notag\\ &\hspace{2.4in}\text{for  } j \in \mK,
\end{align}
where
\begin{align}
&\hspace{-0cm}\Lambda'_{j,\text{LOS}}([0,x))= \notag\\
&2\pi \lambda_j \sum\limits_{d=1}^D \frac{\beta_{jd} \cdot x ^{\frac{2}{\alpha_d^{jL} }-1}}{\alpha_d^{jL} \cdot {\kappa_d^L}^\frac{2}{\alpha_d^{jL}}}\mathds{1}(\kappa_d^L {R_{j(d-1)}^{\alpha_d^{jL}}} <x< \kappa_d^L {R_{jd}^{\alpha_d^{jL}}} ),\\
&\hspace{-0cm} \Lambda'_{j,\text{NLOS}} ([0,x))=  \notag\\
&2\pi \lambda_j \sum\limits_{d=1}^D \frac{(1-\beta_{jd}) \cdot x ^{\frac{2}{\alpha_d^{jN} }-1}}{\alpha_d^{jN}\cdot{\kappa_d^N}^\frac{2}{\alpha_d^{jN}}}\mathds{1}(\kappa_d^N {R_{j(d-1)}^{\alpha_d^{jN}}} <x< \kappa_d^N {R_{jd}^{\alpha_d^{jN}}} ).
\end{align}
\setcounter{TempEqCnt}{\value{equation}}
\setcounter{equation}{32}
\begin{figure*}
 \begin{align}
\hspace{0cm}A_{j,s} =
\begin{cases}
& \frac{\prob_{L_{0,s}}}{\alpha_1^{0s}{\kappa_1^s}^{\frac{2}{\alpha_1^{0s}}}{\sigma^2_j}}  \displaystyle\int_0^{\infty} e^{ \bigg(-\frac{1}{2{\sigma^2_j}}(\frac{l_{0,s}}{\kappa_1^s})^\frac{2}{\alpha_1^{0s}} -\sum_{\substack{k=1}  }^K \Lambda_k([0,\frac{P_k B_k}{P_0 B_0 } l_{0,s})) \bigg) } dl_{0,s}    \hspace{2.1 in} \text{for } j=0,
\\
&\displaystyle \int_0^{\infty} \bigg( \sum\limits_{m \in \{\los, \nlos\}} \prob_{L_{0,m}}  e^{-\frac{1}{2{\sigma^2_j}}(\frac{P_0 B_0 l_{j,s}}{P_j B_j \kappa_1^m})^\frac{2}{\alpha_1^{0,m}} }\bigg)  \Lambda'_{j,s'}([0,l_{j,s})) e^{ \Big(-\sum_{\substack{k=1}  }^K \Lambda_k([0,C_2 = \frac{P_k B_k}{P_j B_j } l_{j,s})) \Big) }  dl_{l,s}    \hspace{0.15in} \text{for } j \in \mK,
 \end{cases}
\end{align}
\begin{align}
\hspace{0cm}A_{j,s} =
\begin{cases}
&\frac{2 \prob_{L_{0,s}}  }{\alpha_1^{ks}{\kappa_1^s}^{\frac{2}{\alpha_1^{ks}}}R^2_j}  \displaystyle\int_0^{\kappa_1^s R_j^ { \alpha_1^{ks} }} l_{0,s}^{{\frac{2}{\alpha_1^{ks}}}-1}  e^ { - \sum_{\substack{k=1}  }^K\Lambda_k([0,\frac{P_k B_k}{P_0 B_0 } l_{0,s}))  }    dl_{0,s} \hspace{2.45 in} \text{for } j=0,
\\
&\displaystyle\int_0^{\infty} \bigg( \sum\limits_{m \in \{\los,\nlos\}} \prob_{L_{0,m}}  \Big( 1-\frac{1}{R^2_j}(\frac{P_0 B_0 l_{j,s}}{P_j B_j \kappa_1^m})^\frac{2}{\alpha_1^{k,m}}  \Big) \bigg)  \Lambda'_{j,s'}([0,l_{j,s})) e^{ \Big(-\sum_{\substack{k=1}  }^K \Lambda_k([0,\frac{P_k B_k}{P_j B_j }  l_{j,s})) \Big) }   dl_{l,s}   \hspace{0.1in} \text{for } j \in \mK,
\end{cases}
\end{align}
 \hrulefill
\end{figure*}
\setcounter{equation}{\value{TempEqCnt}}

\subsection{Association Probability}
In this paper, UEs are assumed to be associated with the BS offering the strongest long-term averaged biased-received power. This can be mathematically expressed as
\begin{align}
P=\mathop{\max}_{j\in \mK_1, i\in \Phi_j} { P_{j,i} B_{j,i} G_0 L_{j,i}^{-1} }
\end{align}
where $P$ is the average biased received power of the typical UE, $P_{j,i},B_{j,i}, L_{j,i}^{-1}$ are the transmission power, biasing factor, and path loss of the $i^{th}$ BS in the $j^{th}$ tier, respectively, and $G_0$ is the effective antenna gain. Since $P_{j,i}$ and $B_{j,i}$ are the same for all BSs in the $j^{th}$ tier, the strongest average biased received power within each tier comes from the BS providing the minimum path loss. Therefore,
\begin{align}
P=\mathop{\max}_{j\in \mK_1} { P_{j} B_{j} G_0 L_{j,min}^{-1} }
\end{align}
where $L_{j,min}$ is the minimum path loss of the typical UE from a BS in the $j^{th}$ tier.

Association probability is defined as the probability that a typical UE is associated with a LOS/NLOS BS in the $j^{th}$ tier for  $j \in \mK_1$, and the result for association probabilities are provided in the following lemma.
\begin{Lemm}
The probability that the typical UE is associated with a LOS/NLOS BS in the $j^{th}$ tier for $ j\in \mK_1$, is
\begin{align}
\hspace{-0.35in}A_{j,s} =
\begin{cases}
 \E_{L_{0,s}} \bigg[ \bigg( \prod\limits_{k=1}^K \overline{F}_{L_k} \big(\frac{P_k B_k}{P_0 B_0 } l_{0,s} \big) \bigg)   \bigg] \prob_{L_{0,s}},   \\ \hspace{6cm} \text{for } j=0,
 \\
\E_{L_{j,s}} \bigg[ \bigg( \overline{F}_{L_0}( \frac{P_0 B_0}{P_j B_j } l_{j,s}) \prod\limits_{\substack{k=1 \\ k\neq j}}^K \overline{F}_{L_{k}}\big(\frac{P_k B_k}{P_j B_j } l_{j,s} \big)  \bigg)  \overline{F}_{L_{j,s'}}(l_{j,s})   \bigg],    \\ \hspace{6cm} \text{for }j\in \mK,
\end{cases}
\end{align}
where $s, s' \in\{ \los, \nlos  \}, s\neq s'$, $\prob_{L_{0,\los}} = \beta_{01}, \prob_{L_{0,\nlos}} =(1- \beta_{01})$, $l_{j,s}$ is the path loss to a LOS/NLOS BS in the $j^{th}$ tier, $ \overline{F}_{L_0}(\cdot)$ is given by (14) or (16) (depending on the cluster process), and $\overline{F}_{L_k}(\cdot)$, and $ \overline{F}_{L_{j,s'}}(\cdot)$ are given by (22) and (26), respectively.

$Proof.$ See Appendix B.
\end{Lemm}

\begin{Corr}
When $\Phi_u^j$ is a Thomas cluster process, the association probability with a LOS/NLOS BS in the $j^{th}$ tier for $j \in \mK_1$, is given in (33) at the top of the page,
where $s\in\{ \text{LOS, NLOS}  \}$, $\Lambda_k([0,\cdot))$ is given in (23),  $\Lambda'_{j,\los}([0,\cdot))$ and $\Lambda'_{j,\nlos}([0,\cdot))$ are given in (24) and (25), respectively.

$Proof.$ See Appendix C.
\end{Corr}

\begin{Corr}
When $\Phi_u^j$ is a Matern cluster process, the association probability with a LOS/NLOS BS in the $j^{th}$ tier for $j \in \mK_1$, is given in (34) at the top of the page.

$Proof.$ See Appendix D.
\end{Corr}

\section{ SINR Coverage Probability Analysis} \label{CP}
In this section, an analytical framework is developed to analyze the downlink SINR coverage probability for a typical UE of $\Phi_u^j$ using stochastic geometry and employing the results obtained in Section III.

\subsection{Signal to Interference Plus Noise Ratio (SINR)}
According to the association policy, a typical UE is served by the BS providing the strongest average biased received power. Therefore, if the typical UE is served by a BS in the $j^{th}$ tier located at a distance $y_j$, there exists no BSs in the $k^{th}$ tier ($\forall k \in \mK_1$), within a disc 
$Q_k$
whose center is the location of the typical UE and the radius is proportional to $\frac{P_kB_k}{P_j B_j}l_{j,s}$. We refer to this disc as the exclusion disc throughout this paper.

If the typical UE is associated with a BS in the $j^{th}$ tier, the interference is due to the BSs lying beyond the exclusion disc. Therefore, the interference from the BSs in the $k^{th}$ tier can be expressed as
\setcounter{equation}{34}
\begin{align}
&I_{j,k} =\sum\limits_{i\in \Phi_k \backslash Q_k}P_k G_{k,i} h_{k,i} L_{k,i}^{-1}
\end{align}
where $P_k$ is the transmit power of the BSs in the $k^{th}$ tier, and $G_{k,i}, h_{k,i}, L_{k,i}$ are the effective antenna gain, the small-scale fading gain and the path loss from the $i^{th}$ BS in the $k^{th}$ tier, respectively. All links are assumed to be subject to independent Rayleigh fading i.e.,  $h_{k,i} \sim \exp(1)$.

\setcounter{TempEqCnt}{\value{equation}}
\setcounter{equation}{40}
\begin{figure*}
\begin{align}
\hspace{-1cm}\prob_{C_{j,s}} =
\begin{cases}
&e^{ -\mu_{0,s} {\sigma^2_{n,0}} } e^{- \sum\limits_{k=1}^K\sum\limits_{G} \sum\limits_{a} \int_{\frac{P_k B_k}{P_0 B_0} l_{0,s}}^{\infty} \Big (1- \frac{1}{(1+ {\mu_{0,s} P_k G l_{k,a}^{-1}} )} \Big) P_{G} \Lambda_{k,a}(dl_{k,a})} \hspace{1.95in} \text{for } j =0,   \\
&\Big( \sum\limits_G \sum\limits_{m} \big(\prob_{L_{0,m}}  P_G \int_{\frac{P_0 B_0}{P_j B_j} l_{j,s}}^{\infty}  \frac{{l_{0,m}}^{\frac{2}{\alpha_1^{0 m}}} e^{-\frac{1}{2{\sigma^2_j}}(\frac{l_{0,m}}{\kappa_1^{m}})^\frac{2}{\alpha_1^{0m}}}}{\alpha_1^{0m}{\kappa_1^{m}}^{\frac{2}{\alpha_1^{0m}}}{\sigma^2_j}( l_{0,m} + \mu_{j,s} P_0  G) } dl_{0,m}  \big)\Big) e^{ -\mu_{j,s} {\sigma^2_{n,j}} } e^{- \sum\limits_{k=1}^K\sum\limits_{G} \sum\limits_{n} \int_{\frac{P_k B_k}{P_j B_j} l_{j,s}}^{\infty} \Big (1- \frac{1}{(1+ {\mu_{j,s} P_k G l_{k,n}^{-1}} )} \Big) P_{G} \Lambda_{k,n}(dl_{k,n})}    \\
&\hspace{ 5.4in}\text{for } j \in \mK,
\end{cases}
\end{align}
\begin{align}
\hspace{-1cm}\prob_{C_{j,s}} =
\begin{cases}
&e^{ -\mu_{0,s} {\sigma^2_{n,0}} } e^{- \sum\limits_{k=1}^K\sum\limits_{G} \sum\limits_{a} \int_{\frac{P_k B_k}{P_0 B_0} l_{0,s}}^{\infty} \Big (1- \frac{1}{(1+ {\mu_{0,s} P_k G l_{k,a}^{-1}} )} \Big) P_{G} \Lambda_{k,a}(dl_{k,a})} \hspace{1.95in} \text{for } j =0,   \\
&\Big( \sum\limits_G \sum\limits_{m} \big(\prob_{L_{0,m}}  P_G  \int_{\frac{P_0 B_0}{P_j B_j} l_{j,s}}^{\kappa_1^m R_j^ { \alpha_1^{km} }}  \frac{2 l_{0,m}^{\frac{2}{\alpha_1^{0m}}}}{\alpha_1^{0m}{\kappa_1^m}^{\frac{2}{\alpha_1^{0m}}}R^2_j ( l_{0,m} + \mu_{j,s} P_0  G) } dl_{0,m} \big)\Big)
e^{ -\mu_{j,s} {\sigma^2 _{n,j}}} e^{- \sum\limits_{k=1}^K\sum\limits_{G} \sum\limits_{n} \int_{\frac{P_k B_k}{P_j B_j} l_{j,s}}^{\infty} \Big (1- \frac{1}{(1+ {\mu_{j,s} P_k G l_{k,n}^{-1}} )} \Big) P_{G} \Lambda_{k,n}(dl_{k,n})}. \\
&\hspace{ 5.4in}\text{for } j \in \mK,
\end{cases}
\end{align}
 \hrulefill
\end{figure*}
\setcounter{equation}{\value{TempEqCnt}}

The SINR experienced at a typical UE associated with a LOS/NLOS BS in the $j^{th}$ tier can expressed as
\begin{align}
\sinr_{j,s} = \frac{P_j G_0 h_{j} L_{j,s}^{-1}}{{\sigma^2_{n,j}} + \sum\limits_{k=0}^K \sum\limits_{i\in \Phi_k \backslash Q_k}P_k G_{k,i} h_{k,i} L_{k,i}^{-1}}
\end{align}
where $s\in \{\los, \nlos \}$, $P_j$ is the transmit power in the $j^{th}$ tier, $G_0$ is the effective antenna gain of the link between the serving BS and the typical UE which is assumed to be $MM$, $\sigma^2_{n,j}$ is the variance of the additive white Gaussian noise component, and $h_{j}$ is the fading gain (i.e., the magnitude-square of the Rayleigh fading coefficient) from the serving BS to the typical UE, i.e., $h_j\sim \exp(1)$.

\subsection{SINR Coverage Probability}
A typical UE is said to be in coverage if the received SINR is larger than a certain threshold $T_j >0$ required for successful reception.
\begin{Def}
Given that the typical UE is associated with a LOS/NLOS BS in the $j^{th}$ tier, the SINR coverage probability of the $j^{th}$ tier is defined as
\begin{align}
\prob_{C_{j,s}} = \prob (\sinr_{j,s}>T_j | t=j)
\end{align}
where t indicates the associated tier and $s \in \{\los, \nlos \}$. Therefore, the total coverage probability of the entire network can be defined as
\begin{align}
\mathds{P} _C = \sum\limits_{j=0}^K \sum\limits_{s \in \{ \los,\nlos \}}  A_{j,s}\mathds{P}_{C_{j,s}}
\end{align}
where $A_{j,s}$ is the association probability of a LOS/NLOS BS in the $j^{th}$ tier, which is given in Lemma 3.
\end{Def}

The exact expressions for the coverage probabilities of each tier are given by the following theorem.
\begin{Lem}
Given that the UE is associated with a LOS/NLOS BS from the $j^{th}$ tier ($j\in \mK_1$), the SINR coverage probabilities are given as
\begin{align}
&\hspace{-0.15in}\prob_{C_{j,s}} = \notag \\
&\hspace{-0.15in} \begin{cases}
e^{ -\mu_{0,s} {\sigma^2_{n,0}} } \prod\limits_{k=1}^K \big(  \cL_{I_{0,k}^{\los}}( \mu_{0,s} )  \cL_{I_{0,k}^{\nlos}}( \mu_{0,s} )  \big)  \hspace{0.5in} (j=0)\\
e^  {-\mu_{j,s} {\sigma^2_{n,j}} } \Big( \sum\limits_m  \prob_{L_{0,m}} \cL_{I_{j,0}^m}( \mu_{j,s}\Big)\prod\limits_{k=1}^K \Big(  \cL_{I_{j,k}^{\los}}( \mu_{j,s} )  \cL_{I_{j,k}^{\nlos}}( \mu_{j,s} )  \Big)  \\  \hspace{2.8in} (j\in \mK),
\end{cases}
\end{align}
where $s\in\{\los, \nlos \}, m\in\{\los, \nlos \} $, $\prob_{L_{0,\los}}= \beta_{01}$, $\prob_{L_{0,\nlos}}=1- \beta_{01}$, $\mu_{j,s}= \frac{  T_j l_{j,s}}{P_j G_0}$, $I_{j,k}^s$ is the interference from the LOS/NLOS BSs in the $k^{th}$ tier to the $j^{th}$ tier, $I_{j,0}$ is the interference form the $0^{th}$ tier to the $j^{th}$ tier, and $\cL_{I_{j,k}^{s}}( \mu_{j,s}) $ is the Laplace transform of $I_{j,k}^{s}$ evaluated at $\mu_{j,s}$.
And the total SINR coverage probability of the $K$-tier heterogeneous mmWave cellular network with user-centric small cell deployment can be obtained as follows:
\begin{align}
\mathds{P} _C= \sum_{s\in\{ \los,\nlos\} } \bigg( A_{0,s}\mathds{P}_{C_{0,s}}+ \sum_{j=1}^{K}A_{j,s} \mathds{P}_{C_{j,s}}  \bigg)
\end{align}
where $A_{0,s},A_{j,s}$ are given in (32).

$Proof.$ See Appendix E.
\end{Lem}

\begin{Corr}
If $\Phi_u^j$ is a Thomas cluster process, the conditional SINR coverage probability $ \prob_{C_{j,s}} $ is given at the top of the next page in (41).

$Proof.$ See Appendix F.
\end{Corr}

\begin{Corr}
If $\Phi_u^j$ is a Mat\'ern cluster process, the conditional SINR coverage probability $ \prob_{C_{j,s}} $ is given at the top of the next page in (42).

$Proof.$ See Appendix F.
\end{Corr}

\section{Numerical Results And Discussions} \label{Num}
In this section, we present several numerical results based on our analyses in Section III and Section IV. Simulation results are also provided to validate the accuracy of our analysis.

In the numerical evaluations and simulations, a 2-tier heterogeneous network model with an additional $0^{th}$ tier, which is the cluster center of the typical UE, is considered. For this 2-tier scenario, $j=1$ and $j=2$ correspond to the picocell and microcell, respectively. In other words, a relatively high-power microcell network coexists with denser but lower-power picocells. UEs are clustered around the BSs in the picocells. Therefore, transmit power of BSs in the $0^{th}$ tier is the same as in the $1^{st}$ tier. For both $1^{st}$ and $2^{nd}$ tiers, $D$-ball approximation is used with $D=2$, while $1$-ball model is employed for the $0^{th}$ tier. Parameter values of this model are listed in Table II.

\begin{table}[htbp]
\caption{Parameter Values Table}
\centering
\begin{tabular}{|l |l |}
\hline
\textbf{Parameters} & \textbf{Values}  \\   \hline
$P_0, P_1, P_2$  &  3dBW, 3dBW,  23dBW \\ \hline
$B_0, B_1, B_2$  &  $1, 1,  1$ \\ \hline
$[R_{11}, R_{12}], [\beta_{11}, \beta_{12}]$ & $[40,60], [1,0]$ \\ \hline
$[R_{21}, R_{22}], [\beta_{21}, \beta_{22}]$ & $[50,200], [0.8,0.2]$ \\ \hline
$[R_{01}], [\beta_{01}]$ & $[40], [1]$ \\ \hline
$ \alpha_d^{j,L}$, $\alpha_d^{j,N}$ $\forall j$, $\forall d$  & $2, 4$ \\  \hline
$\lambda_1, \lambda_2$  & $10^{-4}, 10^{-5}$ \\ \hline
$M, m, \theta$  &  $10dB, -10dB, \pi/6$ \\ \hline
Carrier frequency $(F_c)$  & 28 GHz \\  \hline
$\kappa_d^L=\kappa_d^N$ $\forall d$  & $(F_c / 4 \pi)^2$ \\  \hline
$\sigma^2_{n,j}$ $\forall j$  & $-74dBm$ \\ \hline
\end{tabular}
\end{table}

\begin{figure}
\centering
\begin{minipage} {0.45 \textwidth}
\centering
\includegraphics[width=1\textwidth]{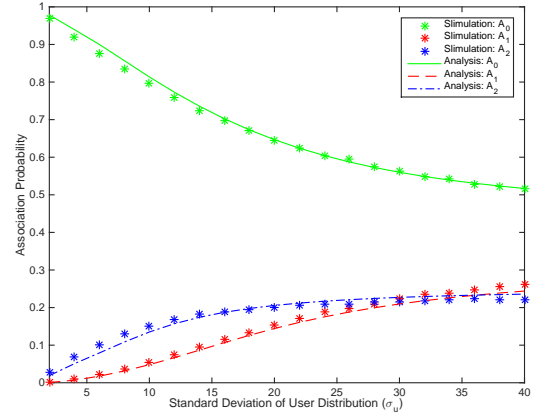} \\
\subcaption{\scriptsize Association probability for the Thomas cluster process. }
\includegraphics[width=1\textwidth]{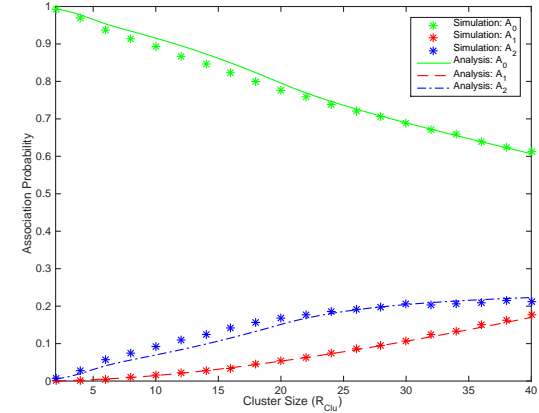}
 \subcaption{\scriptsize Association probability for the Mat\'ern cluster process.}
\end{minipage}
\caption{\small Association probabilities of the two tiers and the cluster center as a function of cluster size, which, for the Thomas cluster process, is given by the standard deviation of UE distribution $\sigma_u$; and for the Mat\'ern cluster process, is given by $R_{clu}$. \normalsize}
\end{figure}

\begin{figure}
\centering
\begin{minipage} {0.45 \textwidth}
\centering
\includegraphics[width=1\textwidth]{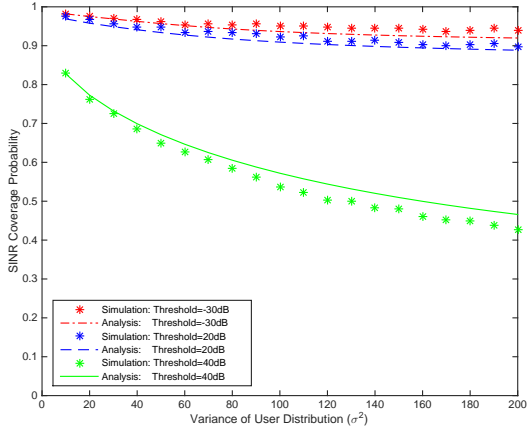} \\
\subcaption{\scriptsize SINR coverage probability for the Thomas cluster process. }
\includegraphics[width=1\textwidth]{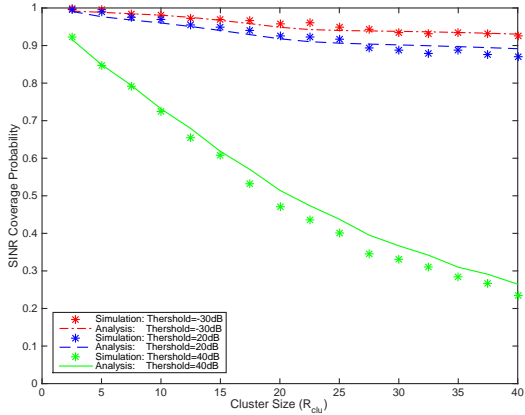}
\subcaption{\scriptsize SINR coverage probability for the Mat\'ern cluster process.}
\end{minipage}
\caption{\small SINR coverage probability for different values of threshold as a function of the: (i) variance of Gaussian UE distribution for the Thomas cluster process and (ii) cluster size for the Mat\'ern cluster process. \normalsize}
\end{figure}

\subsection{Association Probability (AP)}
First, we analyze the effect of UE distribution on the association probability (AP). In Fig. 3, we plot the APs as a function of the cluster size, which is quantified as the standard deviation $\sigma_u$ of Gaussian UE distribution for the Thomas cluster process, and is given by the cluster size $R_{clu}$ of the Mat\'ern cluster process. Since cluster size increases with the increase in $\sigma_u$ and $R_{clu}$, UEs are located relatively farther away from their own cluster center for larger $\sigma_u$ and $R_{clu}$. Therefore, UEs become more likely to connect with the BSs in other picocells and microcells. In other words, AP with the $0^{th}$ tier, $A_0$, decreases, while APs with the $1^{st}$ and $2^{nd}$ tiers, $A_1$ and $A_2$, increases with the increasing cluster size. However, note that UEs are still more likely to associate with the $0^{th}$ tier rather than $1^{st}$ and $2^{nd}$ tiers. We further note that we generally have excellent agreement between simulation and analytical results.

Moreover, we notice in Fig. 3(a) that for the Thomas cluster process, when $\sigma_u$ is less than a certain value (which is approximately $\sigma_u=34$  for this setting), AP with the $1^{st}$ tier is less than that with the $2^{nd}$ tier, while the opposite happens as $\sigma_u$ exceeds 34. Note that with the increase in $\sigma_u$, UEs are more likely to be located farther away from their own cluster center. Since picocell BSs are more densely deployed than microcell BSs, UEs are more likely to be close to another picocell BSs. Thus, $A_1$ becomes greater than $A_2$ for $\sigma_u>34$. However, for the Mat\'ern cluster process, since UEs are uniformly distributed around the cluster center inside a circular disc, UEs cannot be located outside the clusters as shown in Fig. 1(c), and are more compactly distributed. Therefore, $A_2$ is larger than $A_1$ for $R_{clu} < 40$, owing primarily to the larger power in the microcell tier (i.e., the second tier). Note that $P_2 = 23 \text{ dB} > P_1 = 3 \text{ dB}$ as assumed in Table II.



\begin{figure}
\centering
\begin{minipage} {0.45 \textwidth}
\centering
\includegraphics[width=1\textwidth]{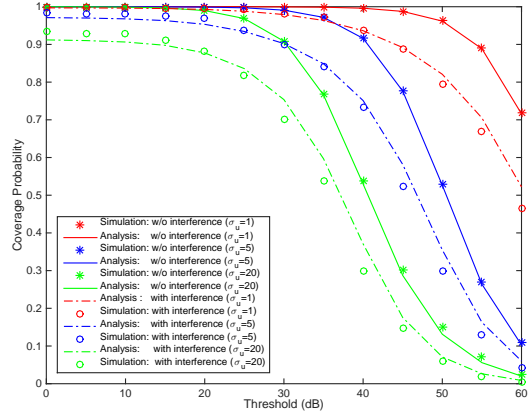} \\
\subcaption{\scriptsize Coverage probability for the Thomas cluster process.}
\includegraphics[width=1\textwidth]{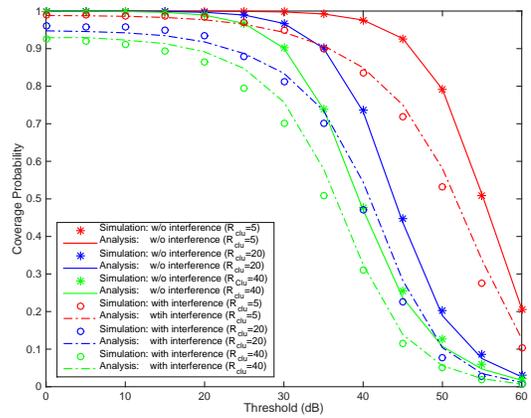}
\subcaption{\scriptsize Coverage probability for the Mat\'ern cluster process.}
\end{minipage}
\caption{\small Comparison of SINR coverage probabilities and SNR coverage probabilities as a function of the threshold in dB for different values of the: (i) standard deviation of UE distribution ($\sigma_u$) for Thomas cluster process or the (ii) cluster size ($R_{clu}$) for Mat\'ern cluster process. \normalsize}
\end{figure}

\subsection{Coverage Probability (CP)}
In this subsection, we investigate the  SINR coverage probability (CP) performance of the network. In Fig. 4, we plot the SINR CP with respect to the variance ($\sigma_u^2$) of UE distribution for the Thomas cluster process and with respect to $R_{clu}$ for the Mat\'ern cluster process. As cluster size increases, we note in both Fig. 4(a) and Fig. 4(b) that SINR CP decreases accordingly. When UEs are close to their cluster center, they are mostly covered by the cluster center (i.e., the $0^{th}$ tier BS). As UEs are distributed far away, probability of being covered by the cluster center goes down accordingly. On the other hand, as shown in Fig. 3, even when $\sigma_u=40$ or $R_{clu} = 40$,  APs of picocells and microcells are small, and thus the probability of being covered by picocells and microcells, other than the $0^{th}$ tier BS, does not increase/improve much. Therefore, as the cluster size increases,  the total SINR CP decreases. Additionally, different curves in Fig. 4(a) and Fig. 4(b) are for different thresholds and we observe that the total SINR CP diminishes with increasing threshold.


In Fig. 5, we plot the total SINR CP and SNR CP as a function of the threshold in dB for different values of standard deviation of UE distribution for Thomas cluster process or the cluster size for Mat\'ern cluster process. In our model, when UE is connected to a picocell or microcell BS outside of its cluster, interference from the $0^{th}$ tier BS at the cluster center is not necessarily negligible due to the relative promixity in the clustered distributions. As expected, relatively large gaps between SINR CP and SNR CP are seen in Fig. 5, indicating that interference has noticeable influence on the CP performance in this clustered system model. We note that this is a departure from mmWave studies with PPP-distributed users, where performance is regarded as noise-limited rather than being interference-limited. On the other hand, different curves in Fig. 5(a) and Fig. 5(b) are for different cluster sizes, and the impact of interference is slightly larger for small-sized clusters.

\begin{figure}
\centering
\begin{minipage} {0.45 \textwidth}
\centering
\includegraphics[width=1\textwidth]{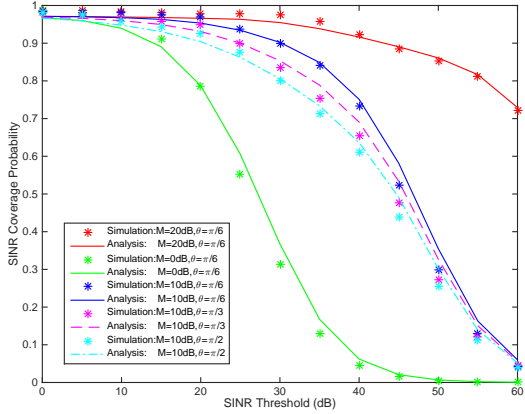} \\
\subcaption{\scriptsize SINR coverage probability for the Thomas cluster process and the standard deviation of UE distribution ($\sigma_u$) is 5.}
\includegraphics[width=1\textwidth]{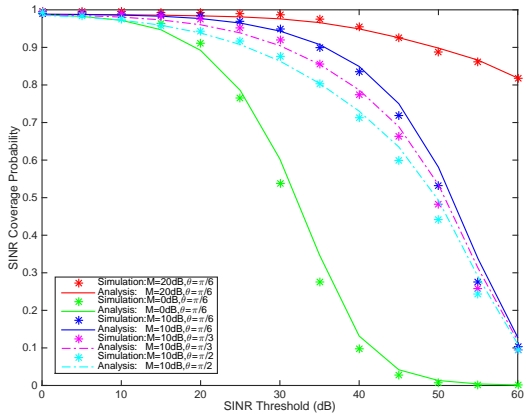}
 \subcaption{\scriptsize SINR coverage probability for the Mat\'ern cluster process and the cluster size ($R_{clu}$) is 5.}
\end{minipage}
\caption{\small SINR coverage probabilities as a function of the threshold in dB for different values of antenna main lobe gain $M$ and the beamwidth of the main lobe $\theta$. \normalsize}
\end{figure}

\begin{figure}
\centering
\includegraphics[width=0.45\textwidth]{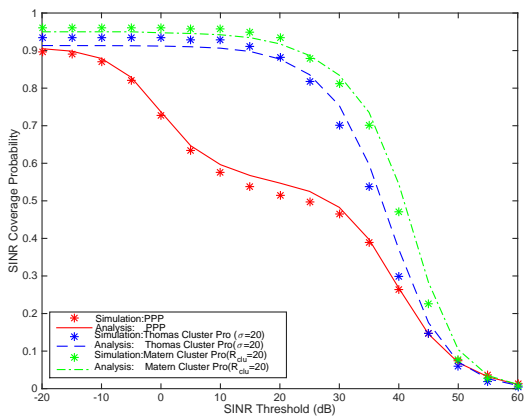}
\caption{SINR coverage probabilities as a function of the threshold in dB when: (i) UE are uniformly distributed and independent of BS locations (PPP); (ii) UE distribution ($\Phi_u^j$) is a Thomas cluster crocess; (iii) UE distribution is a Mat\'ern cluster process.}
\end{figure}

We also investigate the effect of main lobe gain $M$ and different main lobe beamwidth $\theta$ on the SINR coverage probability performance. Improved SINR coverage is achieved when main lobe gain $M$ is increased for the same value of $\theta$ as shown in Fig. 6, since SINR becomes larger with the increase in $M$. On the other hand, for the same $M$, when main lobe beamwidth $\theta$ increases, SINR CP decreases accordingly, as a result of the increase in interference.

Finally, we compare the coverage performances when the UEs are distributed according to PPP or PCP. In Fig. 7, we plot the SINR CP as a function of the threshold. The red solid line represents the scenario in which UEs are uniformly distributed according to a homogeneous PPP and their locations are independent of BS locations. Blue dashed line and green dot-dashed line are for PCP models with UEs being distributed according to a Thomas cluster process and Mat\'ern cluster process, respectively. It is clearly seen that SINR CPs of PCP models are much higher than the SINR CP of the PPP model, indicating that better coverage performance is achieved with user-centric small cell deployments.

\section{Conclusion} \label{Con}
In this paper, we have provided a framework to compute the the SINR CP in a $K$-tier heterogeneous downlink mmWave cellular network with user-centric small cell deployments. A heterogeneous network model is considered, with BSs in each tier being distributed according to PPPs, while UEs being deployed according to a PCP,  i.e., (i) Thomas cluster process, where the UEs are clustered around the base stations (BSs) and the distances between UEs and the BS are modeled as Gaussian distributed, and (ii) Mat\'ern cluster process, where the UEs are scattered according to a uniform distribution. Distinguishing features of mmWave have been incorporated into the analysis, including directional beamforming and a sophisticated path loss model addressing both LOS and NLOS transmissions. In addition, a $D$-ball approximation is applied, to characterize the blockage model, with different path loss exponents being assigned to LOS and NLOS links in different balls. We have determined the CCDF and PDF of the path loss, as well as the association probability of each tier. We have also derived the SINR coverage probability of the entire network using the stochastic geometry framework. Our analysis and numerical results demonstrate that the parameters of the model have significant impact on coverage probability, e.g., CP can be improved, by decreasing the size of UE clusters around BSs, decreasing the beamwidth of the main lobe, or increasing the main lobe directivity gain. Moreover, different from other related works such as \cite{Milli_Esma}, interference in our clustered model has noticeable influence on the coverage performance. Compared with the model in which the UEs are PPP-distributed, our model with user-centric small cell deployments has much larger CP as a function of the SINR threshold. Investigating the interference from which tier has the dominating influence on the performance is considered as future work.

\appendix
\subsection{Proof of Lemma 1}
The CCDF of path loss $L_{0,s} ( \text{ for }s\in \{ \los, \nlos\}$) from the typical UE to a LOS/NLOS BS in the $0^{th}$ tier can be expressed as
\setcounter{equation}{42}
\begin{align}
\overline{F}_{L_{0,s}}(x) &=\prob(l_{0,s} \geq x) \overset{(a)}{=}\prob(\kappa_1^s {y_0}^{\alpha_1^{0s}} \geq x ) \notag \\
&=\prob \Big(y_0 \geq {\big(\frac{x}{\kappa_1^s} \big)}^{\frac{1}{\alpha_1^{0s}}} \Big) \overset{(b)}{=} \overline{F}_{Y_0} \Big( {\big(\frac{x}{\kappa_1^s} \big)}^{\frac{1}{\alpha_1^{0s}}}  \Big)  \notag \\
&\overset{(c)}{=}
\begin{cases}
&\exp(-\frac{1}{2{\sigma^2_j}}(\frac{x}{\kappa_1^s})^\frac{2}{\alpha_1^{0s}}) \quad (x \geq 0 ), \\
& \frac{2{l_{0,s}}^{{\frac{2}{\alpha_1^{ks}}}-1}}{\alpha_1^{ks}{\kappa_1^s}^{\frac{2}{\alpha_1^{ks}}}R^2_j}   \qquad         (0\leq l_{0,s} \leq \kappa_1^s R_j^ { \alpha_1^{ks} })
\end{cases}
\end{align}
where (a) follows from the expression of path loss $L_0$ in (13) on the link in the $0^{th}$ tier, (b) follows from the definition of CCDF,  and (c) is due to the the expression of $\overline{F}_{Y_0}(y_0)$ given in (5) and (7).

Thus, the PDF of path loss $L_{0,s}$ can be obtained as follows:
\begin{align}
f_{L_{0,s}} (x)&=-\frac{d \overline{F}_{L_{0,s}}(x)}{dx} \notag \\
&=
\begin{cases}
& \frac{{x}^{{\frac{2}{\alpha_1^{0s}}}-1}}{\alpha_1^{0s}{\kappa_1^s}^{\frac{2}{\alpha_1^{0s}}}{\sigma^2_j}}\exp(-\frac{1}{2{\sigma^2_j}}(\frac{x}{\kappa_1^s})^\frac{2}{\alpha_1^{0s}})    \quad (x \geq 0 ), \\
& 1-\frac{{l_{0,s}}^{\frac{2}{\alpha_1^{ks}}}}{{\kappa_1^s}^{\frac{2}{\alpha_1^{ks}}}R^2_j}    \qquad         (0\leq l_{0,s} \leq \kappa_1^s R_j^ { \alpha_1^{ks} }) .
\end{cases}
\end{align}

Therefore, the CCDF of the path loss $L_0$ from a typical UE to the BS in the $0^{th}$ tier can be expressed as
\begin{align}
\overline{F}_{L_{0}}(x)
&\overset{(a)}{=} \prob_{L_{0,\los}} \overline{F}_{L_{0,\los}}(x) +\prob_{L_{0,\nlos}} \overline{F}_{L_{0,\nlos}}(x) \notag \\
&= \sum\limits_{s \in \los,\nlos} \prob_{L_{0,s}} \exp(-\frac{1}{2{\sigma^2_j}}(\frac{x}{\kappa_1^s})^\frac{2}{\alpha_1^{0,s}})  (x \geq 0 ),
\end{align}
where (a) follows from the fact that there is only one BS in the $0^{th}$ tier, which could be on a LOS or NLOS link.

The PDF of path loss $L_0$ in (15) can be obtained by differentiating $\overline{F}_{L_{0}}(x)$, with respect to (w.r.t.) $x$.


\subsection{Proof of  Lemma 3}
Note that the association probability of a LOS/NLOS BS in the $j^{th}$ tier is
\begin{align}
\hspace{-.5cm}A_{j,s}
     &\overset{(a)}{=} \mathds {P}\big (\text{typical user is connected to the } j^{th} \text{ tier}\big ) \prob(L_{j,s'} > L_{j,s}) \notag \\
     &= \mathds {P}\big (P_j B_j L_{j,s}^{-1} \geq P_k B_k L_{k}^{-1}, \text{} k\in \mK_1, k\neq j\big)\prob(L_{j,s'} > L_{j,s}) \notag \\
     &= \mathds {P}\left(L_{k} \geq \frac{P_k B_k}{P_j B_j } L_{j,s}, k\in \mK_1, k\neq j\right) \prob(L_{j,s'} > L_{j,s}) \notag \\
     &\overset{(b)}{=}  \prob(L_{j,s'} > L_{j,s}) \prod_{\substack{k=0\\ k\neq j}  }^K \mathds{P}\left(L_{k} \geq \frac{P_k B_k}{P_j B_j } L_{j,s} \right),
\end{align}
where $s' \in \{ \los, \nlos \}$ and $s \neq s'$. (a) follows from the definition of association probability, and (b) is due to the fact that the distributions of $\{L_k\}$ are independent.

\subsubsection{For the $0^{th}$ tier (j=0)}
\begin{align}
A_{0,s} &=\prob(L_{0,s'} > L_{0,s}) \prod_{\substack{k=1}  }^K \mathds{P}\left(L_{k} \geq \frac{P_k B_k}{P_0 B_0 } L_{0,s} \right) \notag \\
&  \overset{(a)}{=}  \prob_{L_{0,s}}  \E_{L_{0,s}} \bigg[ \bigg( \prod_{\substack{k=1}  }^K \overline{F}_{L_k}\big(\frac{P_k B_k}{P_0 B_0 } l_{0,s} \big) \bigg)   \bigg],
\end{align}
where (a) is because of the fact that there is only one BS in the $0^{th}$ tier, and therefore if the BS is on a LOS link, $\prob(L_{0,s'} > L_{0,s})$ can be expressed as $\prob_{L_{0,\los}}$. Also in (a), with the use of expected value w.r.t. $L_{0,s}$, the definition of the CCDF of path loss $L_k$ is applied.


\setcounter{TempEqCnt}{\value{equation}}
\setcounter{equation}{49}
\begin{figure*}
\begin{align}
&A_{j,s}
\overset{(a)}{=} \int_0^{\infty} \bigg[ \bigg( \overline{F}_{L_0} ( C_1 l_{j,s}) \prod_{\substack{k=1 \\ k\neq j}  }^K \overline{F}_{L_{k}}\big(C_2 l_{j,s} \big) \bigg)  \overline{F}_{L_j,s'}(l_{j,s} )   \bigg]  f_{L_{j,s}}(l_{j,s}) dl_{l,s} \notag \\
&\overset{(b)}{=} \int_0^{\infty} \bigg( \sum\limits_{m \in \{ \los,\nlos\}} \prob_{L_{0,m}}  e^{-\frac{1}{2{\sigma^2_j}}(\frac{C_1 l_{j,s}}{\kappa_1^{m}})^\frac{2}{\alpha_1^{0,m}} } \bigg) e^{ \Big(-\sum_{\substack{k=1, k\neq j}  }^K \Lambda_k([0,C_2 l_{j,s})) \Big) } e^{\Lambda_{j,s'}([0,l_{j,s}))} \Lambda'_{j,s'}([0,l_{j,s})) e^{\Lambda_{j,s}([0,l_{j,s}))} dl_{l,s} \notag \\
&\overset{(c)}{=} \int_0^{\infty} \bigg( \sum\limits_{m \in \{ \los,\nlos\}} \prob_{L_{0,m}}  e^{-\frac{1}{2{\sigma^2_j}}(\frac{C_1 l_{j,s}}{\kappa_1^{m}})^\frac{2}{\alpha_1^{0,m}} } \bigg) e^{ \Big(-\sum_{\substack{k=1, k\neq j}  }^K \Lambda_k([0,C_2 l_{j,s})) \Big) } e^{\Lambda_{j}([0,l_{j}))} \Lambda'_{j,s'}([0,l_{j,s}))  dl_{l,s} \notag \\
&\overset{(d)}{=} \int_0^{\infty} \bigg( \sum\limits_{m \in \{ \los,\nlos\}} \prob_{L_{0,m}}  e^{-\frac{1}{2{\sigma^2_j}}(\frac{C_1 l_{j,s}}{\kappa_1^{m}})^\frac{2}{\alpha_1^{0,m}} } \bigg)  \Lambda'_{j,s'}([0,l_{j,s})) e^{ \Big(-\sum_{\substack{k=1}  }^K \Lambda_k([0,C_2 l_{j,s})) \Big) }   dl_{l,s}
\end{align}
 \hrulefill
\end{figure*}
\setcounter{equation}{\value{TempEqCnt}}
\subsubsection{For the $j^{th} $ tier ($j \in \mK$) }
\begin{align}
&\hspace{-0.1in} A_{j,s}
\overset{(a)}{=} \E_{L_{j,s}} \bigg[ \bigg(  \prod_{\substack{k=0 \\ k\neq j}  }^K \overline{F}_{L_{k}}\big(\frac{P_k B_k}{P_j B_j } l_{j,s} \big)  \bigg)  \overline{F}_{L_j,s'}(l_{j,s} )    \bigg]   \notag\\
&\hspace{-0.1in}\overset{(b)}{=} \E_{L_{j,s}}  \bigg[\bigg(  \overline{F}_{L_0}( \frac{P_0 B_0}{P_j B_j } l_{j,s}) \prod_{\substack{k=1 \\ k\neq j}  }^K \overline{F}_{L_{k}}\big(\frac{P_k B_k}{P_j B_j } l_{j,s} \big) \bigg)  \overline{F}_{L_j,s'}(l_{j,s} )   \bigg],  \notag \\
\end{align}
where (a) follows from the definition of the CCDF of path loss $L_k$ and CCDF of path loss $L_{j,s'}$, and by initially considering a fixed value $l_{j,s}$ and then taking the expected value w.r.t. $L_{j,s}$. (b) is due to the fact that the CCDF of $L_0$ is  different from the CCDF of $L_k$, and they should be separately considered.

\subsection{Proof of Corollary 2}
When $\Phi^j_u$ is a Thomas cluster process, the association probability of a LOS/NLOS BS is expressed as follows:
\subsubsection{For the $0^{th}$ tier (j = 0)}
\setcounter{equation}{48}
\begin{align}
&\hspace{-0.in}A_{0,s}
\overset{(a)}{=} \prob_{L_{0,s}}   \int_0^{\infty} \big[ \prod_{\substack{k=1}  }^K \overline{F}_{L_k}\big(\frac{P_k B_k}{P_0 B_0 } l_{0,s} \big)   \big] f_{L_{0,s}}(l_{0,s}) dl_{0,s} \notag \\
&\hspace{-0.in}\overset{(b)}{=} \prob_{L_{0,s}}   \int_0^{\infty} \big[ \prod_{\substack{k=1}  }^K \exp(-\Lambda_k([0,\frac{P_k B_k}{P_0 B_0 } l_{0,s}))   \big] \notag \\  &\hspace{3cm}\frac{{l_{0,s}}^{{\frac{2}{\alpha_1^{0s}}}-1}}{\alpha_1^{0s}{\kappa_1^s}^{\frac{2}{\alpha_1^{0s}}}{\sigma^2_j}}\exp(-\frac{1}{2{\sigma^2_j}}(\frac{l_{0,s}}{\kappa_1^s})^\frac{2}{\alpha_1^{0s}})  dl_{0,s} \notag \\
&\hspace{-0in}=\frac{\prob_{L_{0,s}}}{\alpha_1^{0s}{\kappa_1^s}^{\frac{2}{\alpha_1^{0s}}}{\sigma^2_j}}  \int_0^{\infty} e^{ \bigg(-\frac{1}{2{\sigma^2_j}}(\frac{l_{0,s}}{\kappa_1^s})^\frac{2}{\alpha_1^{0s}}  -\sum_{\substack{k=1}  }^K \Lambda_k([0,\frac{P_k B_k}{P_0 B_0 } l_{0,s})) \bigg) } dl_{0,s},
\end{align}
where (a) follows from the definition of expected value and by plugging in the PDF of $L_{0,s}$, and in (b) the expressions of CCDF of $L_k$ in (16) and PDF of $L_{0,s}$ in (14) are applied.

\subsubsection{For the  $j^{th}$ tier ($j \in \mK$)}
Assume $C_1=  \frac{P_0 B_0}{P_j B_j }, C_2 = \frac{P_k B_k}{P_j B_j } $, then the association probability of the LOS/NLOS BSs in the $j^{th}$ tier is given in (50) at the top of the next page, where $m \in \{ \los, \nlos\}$, and (a) follows from the definition of expected value by plugging in the PDF of path loss $L_{j,s}$, in (b)  the expressions of CCDF of path loss $L_k$ in (16) and the CCDF of $L_0$ in (14) are applied, and (c) is due to the fact that $\Lambda_{j,s}([0,l_{j,s}))+ \Lambda_{j,s'}([0,l_{j,s'})) =\Lambda_{j}([0,l_{j}))$.

\subsection{Proof of Corollary 3}
Similar to the proof of Corollary 2, when $\Phi^j_u$ is a Mat\'ern cluster process, the association probability of a LOS/NLOS BS is expressed as follows for the $0^{th}$ tier and $j^{th}$ tier, respectively:
\setcounter{equation}{50}
\begin{align}
&\hspace{-0.2in}A_{0,s} \notag\\
& \hspace{-0.2in} =\prob_{L_{0,s}}   \int_0^{\kappa_1^s R_j^ { \alpha_1^{ks} }} \Big( \prod_{\substack{k=1}  }^K \exp(-\Lambda_k([0,\frac{P_k B_k}{P_0 B_0 } l_{0,s}))   \Big) \Big( \frac{2{l_{0,s}}^{{\frac{2}{\alpha_1^{ks}}}-1}}{\alpha_1^{ks}{\kappa_1^s}^{\frac{2}{\alpha_1^{ks}}}R^2_j} \Big)  dl_{0,s} \notag\\
&\hspace{-0.2in}= \frac{2 \prob_{L_{0,s}}  }{\alpha_1^{ks}{\kappa_1^s}^{\frac{2}{\alpha_1^{ks}}}R^2_j}  \int_0^{\kappa_1^s R_j^ { \alpha_1^{ks} }} l_{0,s}^{{\frac{2}{\alpha_1^{ks}}}-1}  e^ { - \sum_{\substack{k=1}  }^K\Lambda_k([0,\frac{P_k B_k}{P_0 B_0 } l_{0,s}))  }    dl_{0,s}
\end{align}
\begin{align}
&A_{j,s}=\int_0^{\infty} \bigg( \sum\limits_{m \in \{\los,\nlos\}} \prob_{L_{0,m}}  \Big( 1-\frac{{l_{0,m}}^{\frac{2}{\alpha_1^{km}}}}{{\kappa_1^m}^{\frac{2}{\alpha_1^{km}}}R^2_j}  \Big) \bigg) \notag \\
&\hspace{0.6 in} \Lambda'_{j,s'}([0,l_{j,s})) e^{ \Big(-\sum_{\substack{k=1}  }^K \Lambda_k([0,\frac{P_k B_k}{P_j B_j }  l_{j,s})) \Big) }   dl_{l,s}.
\end{align}

\subsection{Proof of Theorem 1}
Given that the typical UE is associated to a LOS/NLOS BS in the $j^{th}$ tier, the coverage probability can be expressed as
\begin{align}
&\mathds{P}_{C_{j,s}}
= \prob (\sinr_{j,s}>T_j | t=j) \notag \\
&\overset{(a)}{=} \mathds{P}\bigg(\frac{P_j  G_0 h_j l_{j,s}^{-1}}{{\sigma^2_{n,j}} + \sum\limits_{k=0}^K I_{j,k} }> T_j \bigg) \notag \\
&=\mathds{P}\left(h_{j}> \frac{T_j l_{j,s}}{P_j G_0} \left({\sigma^2_{n,j}} + \sum\limits_{k=0}^K I_{j,k}\right)\right)
\notag \\
&\overset{(b)}{=}\E \exp \bigg( -\frac{  T_j l_{j,s}}{P_j G_0} ({\sigma^2_{n,j}} + \sum\limits_{k=0}^K I_{j,k})     \bigg) \notag \\
&\overset{(c)}{=} e^{-\mu_{j,s} {\sigma^2_{n,j}} } \E \exp \Big(  \sum\limits_{k=0}^K I_{j,k} \Big) \notag \\
&\hspace{0cm}\overset{(d)}{=}
\begin{cases}
e^{ -\mu_{0,s} {\sigma^2_{n,0}} } \prod\limits_{k=1}^K  \cL_{I_{0,k}}( \mu_{0,s} )     \qquad \qquad  (j=0)\\
e^  {-\mu_{j,s} {\sigma^2_{n,j}} } \cL_{I_{j,0}}( \mu_{j,s} ) \prod\limits_{k=1}^K  \cL_{I_{j,k}}( \mu_{j,s} )   \quad (j\in \mK)
\end{cases} \notag \\
&\hspace{0.in}\overset{(e)}{=}
\begin{cases}
e^{ -\mu_{0,s} {\sigma^2_{n,0}} } \prod\limits_{k=1}^K \big(  \cL_{I_{0,k}^{\los}}( \mu_{0,s} )  \cL_{I_{0,k}^{\nlos}}( \mu_{0,s} )  \big)  \hspace{0.5in} (j=0)\\
e^  {-\mu_{j,s} {\sigma^2_{n,j}} } \Big( \sum_m  \prob_{L_{0,m}} \cL_{I_{j,0}^m}( \mu_{j,s}\Big)\prod\limits_{k=1}^K \Big(  \cL_{I_{j,k}^{\los}}( \mu_{j,s} )  \cL_{I_{j,k}^{\nlos}}( \mu_{j,s} )  \Big)  \\  \hspace{2.8in} (j\in \mK),
\end{cases}
\end{align}
where (a) follows from the fact that if a given typical UE is associated to the $j^{th}$ tier, then $\sinr_{j,s}=\frac{P_j  G_0 h_j l_{j,s}^{-1}}{{\sigma_j}^2 + \sum\limits_{k=0}^K I_{j,k} }$. (b) follows from $h_j \sim \exp(1)$. (c) is due to the independence of noise and interference terms. (d) follows from the fact that for the $0^{th}$ tier, interference links come from all $K$ tiers, while for the $j^{th}$ tier $(j \in \mK)$, interference links come from all $K$ tiers and the $0^{th}$ tier. (e) is because for the $0^{th}$ tier, only one BS exists, so that
\begin{align}
\cL_{I_{j,0}}( \mu_{j,s} ) &= \prob_{L_{0,\los}} \cL_{I_{j,0}^{\los}}( \mu_{j,s} ) + \prob_{L_{0,\nlos}} \cL_{I_{j,0}^{\nlos}}( \mu_{j,s} ) \notag \\
&=\sum\limits_{m\in \{ \los, \nlos\}} \prob_{L_{0,m}} \cL_{I_{j,0}^m}( \mu_{j,s} ),
\end{align}
and for the $j^{th}$ tier, both LOS links and NLOS links exist and they are independent, so that
\begin{align}
\cL_{I_{j,k}}( \mu_{j,s} ) = \cL_{I_{j,k}^{\los}}( \mu_{j,s}) \cL_{I_{j,k}^{\nlos}}( \mu_{j,s}) .
\end{align}

\subsection{Proof of Corollaries 4 and 5}
\subsubsection{Interference from the $k^{th}$ tier ($k\in \mK$)}
When effective antenna gain $G\in\{MM, Mm, mm\}$ is considered, tools from stochastic geometry can be applied to compute the Laplace transforms of interference from the $k^{th}$ tier ($I_{j,k}^s$), which can be split into three parts
\begin{align}
I_{j,k}^s = I_{j,k}^{s,MM} + I_{j,k}^{s,Mm} +I_{j,k}^{s,mm} = \sum\limits_{G \in \{ MM, Mm, mm\}} I_{j,k}^{s,G},
\end{align}
where $I_{j,k}^{s,G}$ denotes the interference with random effective antenna gain. In addition, according to the thinning theorem, each independent PPP has a density of $\lambda_j P_G$ \cite{Milli_Esma}, where $P_G$ is given in (11).

Hence, Laplace transform of the interference from the $k^{th}$ tier can be expressed as
\begin{align}
\cL_{I_{j,k}^s }(u) &=\E \exp(-\mu_{j,k} {I_{j,k}^s } ) = \E \exp(-\mu_{j,k} \sum\limits_{G} I_{j,k}^{s,G} )  \notag\\
&=\prod_{G} \E \exp(-\mu_{j,k} {I_{j,k}^{s,G} })=\prod_{G} \cL_{I_{j,k}^{s,G}}(\mu_{j,k}),
\end{align}
where $G \in \{ MM, Mm, mm\}$.

Using the same approach as in \cite{Milli_Esma} (Equation (40), Appendix C), $\E \exp(-\mu_{j,k} {I_{j,k}^{s,G} })$ can be expressed as follows:
\begin{align}
&\E \exp(-\mu_{j,s} I_{j,k}^{\los,G}) \notag \\
&=e^{-\int_{\frac{P_k B_k}{P_j B_j} l_{j,s}}^{\infty} \Big (1- \frac{1}{(1+ {\mu_{j,s} P_k G l_{k,\los}^{-1}} )} \Big) P_{G} \Lambda_{k,\los}(dl_{k,\los})  }, \\
&\E \exp(-\mu_{j,s} I_{j,k}^{\nlos,G}) \notag \\
&=e^{-\int_{\frac{P_k B_k}{P_j B_j} l_{j,s}}^{\infty}  \Big(1- \frac{1}{(1+{\mu_{j,s} P_k G l_{k,\nlos}^{-1}} )} \Big)P_{G} \Lambda_{k,\nlos}(dl_{k,\nlos})  }.
\end{align}

\subsubsection{Interference from the $0^{th}$ tier, (k=0)}
Since there is only one BS in the $0^{th}$ tier and effective antenna gain $G\in\{MM, Mm, mm\}$ is considered, Laplace transform of interference from the $0^{th}$ tier $\cL_{I_{j,0}^s }(\mu_{j,s})$ can be obtained as
\begin{align}
&\cL_{I_{j,0}^s }(\mu_{j,s})=\E \exp(-\mu_{j,s} {I_{j,0}^s } ) =\E_G \big[ \E \exp(-\mu_{j,s} {I_{j,0}^{s,G} } )   \big] \notag \\
&=\sum_{G\in\{MM, Mm, mm\}} P_G \E \exp(-\mu_{j,s} {I_{j,0}^{s,G} }) =\sum_{G} P_G \cL_{I_{j,0}^{s,G} }(\mu_{j,s}).
\end{align}
Additionally, $\E \exp(-\mu_{j,k} {I_{j,0}^{s,G} })$ can be expressed as follows:
\begin{align}
&\E \exp(-\mu_{j,s} I_{j,0}^{\los,G}) \notag \\
&\overset{(a)}{=} {\E_{L_{0,\los}} \Big[\E_{h_0} [\exp(-\mu_{j,s} P_0 h_0 G l_{0,\los}^{-1})]      \Big] } \notag \\
&\overset{(b)}{=} {\E_{L_{0,\los}} \Big[  \frac{1}{(1+ {\mu_{j,s}  P_0 G l_{0,\los}^{-1}})}  \Big] }\notag\\
&\overset{(c)}{=}   \int_{\frac{P_0 B_0}{P_j B_j} l_{j,s}}^{\infty}  \frac{1}{(1+ {\mu_{j,s} P_0  G l_{0,\los}^{-1}}) }  f_{L_{0,\los}}(l_{0,\los}  )dl_{0,\los}  \notag \\
&\overset{(d)}{=}
\begin{cases}
&\displaystyle\int_{\frac{P_0 B_0}{P_j B_j} l_{j,s}}^{\infty}  \frac{l_{0,\los}^{\frac{2}{\alpha_1^{0L}}} e^{-\frac{1}{2{\sigma^2_j}}(\frac{l_{0,\los}}{\kappa_1^L})^\frac{2}{\alpha_1^{0L}}}}{\alpha_1^{0L}{\kappa_1^L}^{\frac{2}{\alpha_1^{0L}}}{\sigma^2_j}( l_{0,\los} + \mu_{j,s} P_0  G) } dl_{0,\los} \\
& \qquad \qquad \qquad \quad \text{if $\Phi^i_u$ is a Thomas cluster process}; \\
&\displaystyle\int_{\frac{P_0 B_0}{P_j B_j} l_{j,s}}^{\kappa_1^L R_j^ { \alpha_1^{kL} }}  \frac{2 l_{0,\los}^{\frac{2}{\alpha_1^{0L}}}}{\alpha_1^{0L}{\kappa_1^L}^{\frac{2}{\alpha_1^{0L}}}R^2_j ( l_{0,\los} + \mu_{j,s} P_0  G) } dl_{0,\los}  \\
& \qquad \qquad \qquad \quad \text{if $\Phi^i_u$ is a Mat\'ern cluster process}; \\
\end{cases}
\end{align}
where, (a) follows from the expression of $I_{j,0}^{s,G}$, (b) is due to $h_0 \sim \exp(1)$, (c) follows from the definition of expected value w.r.t. $L_{0,\los}$, by plugging in the PDF of $L_{0,\los}$, and in (d) the expression of $ f_{L_{0,\los}}$ in (19) and (21) are applied, depending on the cluster process.

With the same method, we can get
\begin{align}
&\hspace{-3.3in}\E \exp \Big (-\mu_{j,s}   I_{j,0}^{\nlos,G} \Big) =  \notag \\
\begin{cases}
& \displaystyle \int_{\frac{P_0 B_0}{P_j B_j} l_{j,s}}^{\infty}  \frac{{l_{0,\nlos}}^{\frac{2}{\alpha_1^{0N}}} e^{-\frac{1}{2{\sigma^2_j}}(\frac{l_{0,\nlos}}{\kappa_1^N})^\frac{2}{\alpha_1^{0N}}}}{\alpha_1^{0N}{\kappa_1^N}^{\frac{2}{\alpha_1^{0N}}}{\sigma^2_j}( l_{0,\nlos} + \mu_{j,s} P_0  G) } dl_{0,\nlos} \\
& \qquad \qquad \qquad \quad \text{if $\Phi^i_u$ is a Thomas cluster process}; \\
&\displaystyle\int_{\frac{P_0 B_0}{P_j B_j} l_{j,s}}^{\kappa_1^N R_j^ { \alpha_1^{kN} }}  \frac{2 l_{0,\nlos}^{\frac{2}{\alpha_1^{0N}}}}{\alpha_1^{0N}{\kappa_1^N}^{\frac{2}{\alpha_1^{0N}}}R^2_j ( l_{0,\nlos} + \mu_{j,s} P_0  G) } dl_{0,\nlos}  \\
& \qquad \qquad \qquad \quad \text{if $\Phi^i_u$ is a Mat\'ern cluster process}. \\
\end{cases}
\end{align}

\setcounter{TempEqCnt}{\value{equation}}
\setcounter{equation}{63}
\begin{figure*}
\begin{align}
&P_{C_{j,s}}(T) 
=e^  {-\mu_{j,s}\sigma^2_{n,j} }\Big( \sum_G \sum_{m} \big(\prob_{L_{0,m}}  P_G \cL_{I_{j,0}^{m,G}}( \mu_{j,s} )  \big)\Big) \prod\limits_{k=1}^K \prod\limits_{G}  \big(  \cL_{I_{j,k}^{\los,G}}( \mu_{j,s} )  \cL_{I_{j,k}^{\nlos,G}}( \mu_{j,s} )  \big) \notag \\
&=
\begin{cases}
&\Big( \sum\limits_G \sum\limits_{m} \big(\prob_{L_{0,m}}  P_G \int_{\frac{P_0 B_0}{P_j B_j} l_{j,s}}^{\infty}  \frac{{l_{0,m}}^{\frac{2}{\alpha_1^{0 m}}} e^{-\frac{1}{2{\sigma^2_j}}(\frac{l_{0,m}}{\kappa_1^{m}})^\frac{2}{\alpha_1^{0m}}}}{\alpha_1^{0m}{\kappa_1^{m}}^{\frac{2}{\alpha_1^{0m}}}{\sigma^2_j}( l_{0,m} + \mu_{j,s} P_0  G) } dl_{0,m}  \big)\Big)
e^{ -\mu_{j,s} {\sigma^2 _{n,j}}} e^{- \sum\limits_{k=1}^K\sum\limits_{G} \sum\limits_{n} \int_{\frac{P_k B_k}{P_j B_j} l_{j,s}}^{\infty} \Big (1- \frac{1}{(1+ {\mu_{j,s} P_k G l_{k,n}^{-1}} )} \Big) P_{G} \Lambda_{k,n}(dl_{k,n})}, \\
& \hspace{4.5in} \text{if $\Phi^i_u$ is a Thomas cluster process}; \\
&\Big( \sum\limits_G \sum\limits_{m} \big(\prob_{L_{0,m}}  P_G  \int_{\frac{P_0 B_0}{P_j B_j} l_{j,s}}^{\kappa_1^m R_j^ { \alpha_1^{km} }}  \frac{2 l_{0,m}^{\frac{2}{\alpha_1^{0m}}}}{\alpha_1^{0m}{\kappa_1^m}^{\frac{2}{\alpha_1^{0m}}}R^2_j ( l_{0,m} + \mu_{j,s} P_0  G) } dl_{0,m} \big)\Big)
e^{ -\mu_{j,s} {\sigma^2 _{n,j}}} e^{- \sum\limits_{k=1}^K\sum\limits_{G} \sum\limits_{n} \int_{\frac{P_k B_k}{P_j B_j} l_{j,s}}^{\infty} \Big (1- \frac{1}{(1+ {\mu_{j,s} P_k G l_{k,n}^{-1}} )} \Big) P_{G} \Lambda_{k,n}(dl_{k,n})}, \\
& \hspace{4.5in} \text{if $\Phi^i_u$ is a Mat\'ern cluster process}.
\end{cases}
\end{align}
 \hrulefill
\end{figure*}
\setcounter{equation}{\value{TempEqCnt}}

Finally, considering $\Phi_i $ is either a Thomas cluster process or a Mat\'ern cluster process, and by combining (53), (57), (58) and (59), we can express the coverage probability of the $0^{th}$ tier as
\begin{align}
&\hspace{-0.in}P_{C_{0,s}}(T)= e^{ -\mu_{0,s} {\sigma^2_{n,0}} } \prod\limits_{k=1}^K   \prod\limits_{G}  \big(  \cL_{I_{0,k}^{\los,G}}( \mu_{0,s} )  \cL_{I_{0,k}^{\nlos,G}}( \mu_{0,s} ) \big) \notag \\
&\hspace{-0.in}=e^{ -\mu_{0,s} {\sigma^2_{n,0}} } \prod\limits_{k=1}^K\prod\limits_{G}  e^{-\int_{\frac{P_k B_k}{P_0 B_0} l_{0,s}}^{\infty} \Big (1- \frac{1}{(1+ {\mu_{0,s} P_k G l_{k,\los}^{-1}} )} \Big) P_{G} \Lambda_{k,\los}(dl_{k,\los})  }  \notag \\
& \hspace{0.1in}\times e^{-\int_{\frac{P_k B_k}{P_0 B_0} l_{0,s}}^{\infty}  \Big(1- \frac{1}{(1+{\mu_{0,s} P_k G l_{k,\nlos}^{-1}} )} \Big)P_{G} \Lambda_{k,\nlos}(dl_{k,\nlos})  } \notag \\
&\hspace{-0.in}=e^{ -\mu_{0,s} {\sigma^2_{n,0}} } e^{ - \sum\limits_{k=1}^K\sum\limits_{G} \sum\limits_{a} \int_{\frac{P_k B_k}{P_0 B_0} l_{0,s}}^{\infty} \Big (1- \frac{1}{(1+ {\mu_{0,s} P_k G l_{k,a}^{-1}} )} \Big) P_{G} \Lambda_{k,a}(dl_{k,a})} .
\end{align}

By combining (53), (57), (58), (59), (60), (61) and (62), the equation of coverage probability of LOS/NLOS BSs of the $k^{th}$ tier $(k \in \mK)$, can be obtained as in (64) at the top of the next page.

\bibliographystyle{ieeetr}
\bibliography{v14}

\end{document}